\documentclass[aps,prc,letterpaper,11pt,twoside,tightenlines,nofootinbib,showpacs,preprint]{revtex4-1}
\usepackage{graphicx}
\usepackage[sort&compress]{natbib}
\usepackage{subfigure}
\usepackage{amsmath}
\usepackage{amsfonts}
\usepackage{cancel}
\usepackage{amssymb}
\usepackage{hyperref}
\usepackage{color}

\begin{document}
\arraycolsep1.5pt
\newcommand{\Ima}{\textrm{Im}}
\newcommand{\Rea}{\textrm{Re}}
\newcommand{\mev}{\textrm{ MeV}}
\newcommand{\be}{\begin{equation}}
\newcommand{\ee}{\end{equation}}
\newcommand{\ba}{\begin{eqnarray}}
\newcommand{\ea}{\end{eqnarray}}
\newcommand{\gev}{\textrm{ GeV}}
\newcommand{\nn}{{\nonumber}}
\newcommand{\dtres}{d^{\hspace{0.1mm} 3}\hspace{-0.5mm}}
\newcommand{\rts}{ \sqrt s}
\newcommand{\non}{\nonumber \\[2mm]}
\raggedbottom

\title{Prediction of an $I=1$  $D \bar D^*$ state and relationship to the claimed $Z_c(3900)$, $Z_c(3885)$.}

\author{F. Aceti$^{1}$,  M. Bayar$^{2}$, E. Oset$^{1}$, A. Mart\'inez Torres$^{3}$, K. P. Khemchandani$^{3}$, J. M. Dias$^{1,\,3}$, F. S. Navarra$^{3}$, M. Nielsen$^{3}$}
\affiliation{$^{1}$Departamento de F\'{\i}sica Te\'orica, Universidad de Valencia and IFIC, Centro Mixto Universidad de
Valencia-CSIC, Institutos de Investigaci\'on de Paterna, Aptdo. 22085, 46071 Valencia,
Spain\\
\\$^{2}$Department of Physics, Kocaeli University, 41380 Izmit, Turkey\\
\\$^{3}$Instituto de F\'isica, Universidade de S\~ao Paulo,
C.P. 66318, 05389-970 S\~ao Paulo, SP, Brazil\\
 }

\date{\today}

\begin{abstract}
 We study here the interaction of $D \bar D^*$ in the isospin $I=1$ channel in the light of recent theoretical advances that allow to combine elements of the local hidden gauge approach with heavy quark spin symmetry. We find that the exchange of light $q \bar q$ is OZI suppressed and, thus, we concentrate on the exchange of heavy vectors and of two pion exchange. The latter is found to be small compared to the exchange of heavy vectors, which then determines the strength of the interaction. A barely $D\bar{D}^*$ bound state decaying into $\eta_c\rho$ and $\pi J/\psi$ is found. At the same time we reanalyse the data of the BESIII experiment on  $e^+ e^- \to \pi^{\pm} (D \bar D^*)^\mp$, from where a $Z_c(3885)$ state was claimed, associated to a peak in the $(D \bar D^*)^\mp$ invariant mass distribution close to threshold, and we find the data compatible with a resonance with mass around $3875$ MeV and width around $30$ MeV. We discuss the possibility that this and the $Z_c(3900)$ state found at BESIII, reconfirmed at $3894$ MeV at Belle, or $3885$ MeV at CLEO, could all be the same state and correspond to the one that we find theoretically.  
\end{abstract}

\maketitle

\section{Introduction}
\label{Intro}

The interaction of mesons with opposite charm to give hidden charm heavy mesons is capturing much attention recently. Indeed, the large number of X, Y, Z states being reported  experimentally \cite{Ali:2011vy,Gersabeck:2012rp,Olsen:2012zz,Li:2012pd} are finding difficulties to be fitted in the ordinary order of standard charmonium states \cite{Brambilla:2010cs} and call for more complex structures. The molecular picture of states coming from the interaction of $D$ or $\bar D^*$ has been one of the sources to interpret some of these states, and different combinations of such mesons giving hidden charm mesons have been considered. In this sense, a bound state of $D \bar D$ was theoretically found in \cite{daniddbar} and tentatively called X(3700). Other works have also reported on this possibility \cite{HidalgoDuque:2012pq,Nieves:2012tt,Guo:2013sya,Liu:2010xh,Zhang:2006ix,zzy1}. Subsequently, experimental support for such a state was found in \cite{danibump} from a bump close to the threshold of the  $D \bar D$  invariant mass distribution in the $e^+e^− \to J/\psi D \bar D $ reaction \cite{Abe:2007sya}. 

The $D^* \bar D^*$ interaction has also been studied \cite{raquelxyz,HidalgoDuque:2012pq}. In \cite{raquelxyz} an extension of the interaction from the local hidden gauge approach \cite{hidden1,hidden2,hidden3,hidden4} was used and several states in different spin-isospin channels were found, some of which could be associated to known X,Y,Z states. The isospin $I=1$ states are more difficult to obtain within this approach since the interaction is weaker in this channel. Even then, a state with $I=1$ and $J=2$ was found in \cite{raquelxyz}, prior to the reports of the $I=1$ $Z_c(4020)$ \cite{Ablikim:2013wzq} found in the $e^+ e^- \to \pi^+ \pi^- h_c$ reaction looking at the invariant mass of $\pi^{\pm} h_c$, or the claimed $Z_c(4025)$ from a peak in the $(D^* \bar D^*)^{\pm}$ spectrum close to threshold
\cite{besexp}. The interpretation of this peak as a $J^P=1^+$ new state with mass 4025 MeV has been scrutinized in \cite{alberdd} where it was found that the peak seen was compatible with a 
$J^P=2^+$ state with mass around 3990 MeV and a width around 160 MeV. Subsequently, the analysis of \cite{raquelxyz} has been revised in \cite{melafran} in the light of the heavy quark spin symmetry (HQSS) and it was found that the binding is smaller than found in 
\cite{raquelxyz}, compatible with the mass suggested in \cite{alberdd} and with a similar width.  

 The $D \bar D^*$ systems have been the most studied, stimulated by the large impact that the X(3872) state \cite{Acosta:2003zx} has had in this field \cite{Swanson:2006st,daniel,Liu:2008tn,Dong:2009yp,Lee:2009hy,FernandezCarames:2009zz,
Matheus:2009vq,Ortega:2010qq,Coito:2010if,Prelovsek:2013cra,Takizawa:2012hy}. Much at the origin, this state was assumed to be a $D^0 \bar D^{*0}$ \cite{Braaten:2003he,Close:2003sg}, however, subsequent works have stressed the relevance of considering the charged component $D^+ \bar D^{*-}$ forming a quite good isospin I=0 state 
\cite{Gamermann:2009fv,Gamermann:2009uq,Dong:2009yp}. More recently, the radiative decay of the X(3872) into $\gamma J/\psi$ has shown that the charged components are essential to obtain the right rates \cite{Nielsen:2010ij,Dong:2008gb,Aceti:2012cb}. Once again, it was surprising to find $I=1$ states, since the interaction in this channel is weaker than for I=0. Yet, experimental work has been conducted recently and the BESIII collaboration has reported a state $Z_c(3900)$ from the invariant mass of $\pi J/\psi$ in the  $e^+ e^- \to \pi^+ \pi^- J/\psi$ reaction 
\cite{Ablikim:2013mio}, with a width of $46\pm 10 \pm 20$ MeV.  The Belle Collaboration has reconfirmed the finding and, using different energies for the electron beam, a peak is also seen in $\pi J/\psi$ around 3894 MeV and a width of about $63\pm 24 \pm 26$ MeV \cite{Liu:2013dau}. CLEO has followed with more precision and reported a peak at 3886 MeV and a width of $37 \pm 4 \pm 8$ MeV \cite{Xiao:2013iha}.  The state observed has $I=1$ and $J^P=1^+$.
   
Theoretical work has followed: in \cite{Voloshin:2013dpa} a discussion is made on possible structures of this state and suggestions of new experiments are made to get a further insight on its nature. A $D \bar D^*$ molecular structure is suggested in \cite{Wilbring:2013cha,Wang:2013hga,Dong:2013iqa,Ke:2013gia}. Work has also been done using QCD sum rules, suggesting a tetraquark structure. In particular, in \cite{Dias:2013xfa} a tetraquark interpolating current was used in order to estimate the decay width of the $Z_c(3900)$, while in  \cite{Qiao} the same tetraquark current is used to estimate the mass.

  In the present work we use an extrapolation of the chiral symmetry approach for the pseudoscalar-vector interaction used in \cite{Birse:1996hd,luisaxial}. This approach was extrapolated to the charm sector in \cite{daniel}, where several axial vector states were obtained from the interaction, among them the X(3872). Yet, in \cite{daniel} no states in $I=1$ for $D \bar D^*$ were found, the interaction being  weaker in this channel than in I=0. Meanwhile, several works have shown the relevance of heavy quark spin symmetry (HQSS) in dealing with the interaction of heavy mesons and how the dynamics of the local hidden gauge approach provides a natural extension of chiral symmetry to the heavy sector, since it respects the rules of HQSS for the dominant terms that come from the exchange of light vectors \cite{xiaojuan,xiaoyo,xiaoaltug}. Further clarifying is the work of \cite{xiaoliang}, where the impulse approximation is used at the quark level to provide an easy interpretation of the HQSS, showing then how to extrapolate  the local hidden gauge approach to the heavy quark sector. All these ideas have been put together in \cite{melafran} to study the $D^* \bar D^*$ interaction in $I=1$. In that work it is shown how the exchange of a light $q \bar q$ is OZI forbidden in $I=1$, which makes the combined exchange of  the SU(3) nonet of pseudoscalar cancel in the limit of equal masses, and the exchange of $\rho, \omega$ also cancel. As a consequence, only the $J/\psi$ exchange is allowed in the case of $I=1$, plus the simultaneous two pion exchange, which was evaluated in  \cite{melafran} but found weaker than the exchange of the vector meson. In spite of the large mass of the $J/\psi$, which suppresses the propagator in the $J/\psi$ exchange, it was found in  \cite{melafran} that the interaction could bind the $D^* \bar D^*$ system weakly and at the same time provide an explanation for the experimental peak in the $D^* \bar D^*$ mass distribution from where the $Z_c(4025)$ was claimed \cite{besexp}. One reason why a weak state not seen before is now obtained has to be found in the improvements on the interaction in the light of HQSS and on the extended range of the momenta allowed in the intermediate states, since the small mass of the light vectors restricts the momenta in the loops to a much larger extent than the exchange of heavy vectors. 

   Another aspect that one should take into consideration is the fact that, similarly to the case of the claimed $Z_c(4025)$ from the peak in the invariant mass of the $(D^* \bar D^*)^{\pm}$ close to threshold, in this case there is also another reaction,  $e^+ e^- \to \pi^{\pm} (D \bar D^*)^\mp$  measured at BESIII \cite{Ablikim:2013xfr}, where the peak in the $(D \bar D^*)^\mp$ invariant mass is interpreted in terms of a new $J^P=1^+$ resonance with mass around 3885 MeV and width $25 \pm 3 \pm 11$ MeV.  It is unclear whether this state is the same as the one claimed in BESIII \cite{Ablikim:2013mio}, or Belle \cite{Liu:2013dau} or CLEO 
\cite{Xiao:2013iha}.  In view of the present situation we combine in this paper the two lines of work in \cite{alberdd} and \cite{melafran} and perform a theoretical study of the  $D \bar D^*$ interaction with the extended hidden gauge approach.  After this, we perform an empirical analysis of the data from the $e^+ e^- \to \pi^{\pm} (D \bar D^*)^\mp$ reaction and see if they can be interpreted in terms of the theoretically found resonance.  The answer to the question is yes and we propose to interpret the data in terms of a resonance $Z_c$ with a mass around $3875$ MeV and a width around $30$ MeV, coming from the $\eta_c\rho$ and $\pi J/\psi$ decay channels. 

In this work we study the $D\bar{D}^*$ system taking into account the possible sources of interaction in order to compare them and identify the most relevant process. We start analysing the contribution coming from the exchange of heavy vectors, proceeding then to the evaluation of the exchange of one light pseudoscalar ($\pi$, $\eta$, $\eta^{\prime}$), followed by the exchange of two correlated and also uncorrelated pions. We find the last three processes very small compared to the heavy vector exchange, which, as in the case of Ref. \cite{melafran}, is found to be the leading source of interaction and, even if small, it is enough to bind the system.
%**************************************************************************************************

\section{Formalism}
\begin{figure}[htpb]
\centering
\includegraphics[scale=0.45]{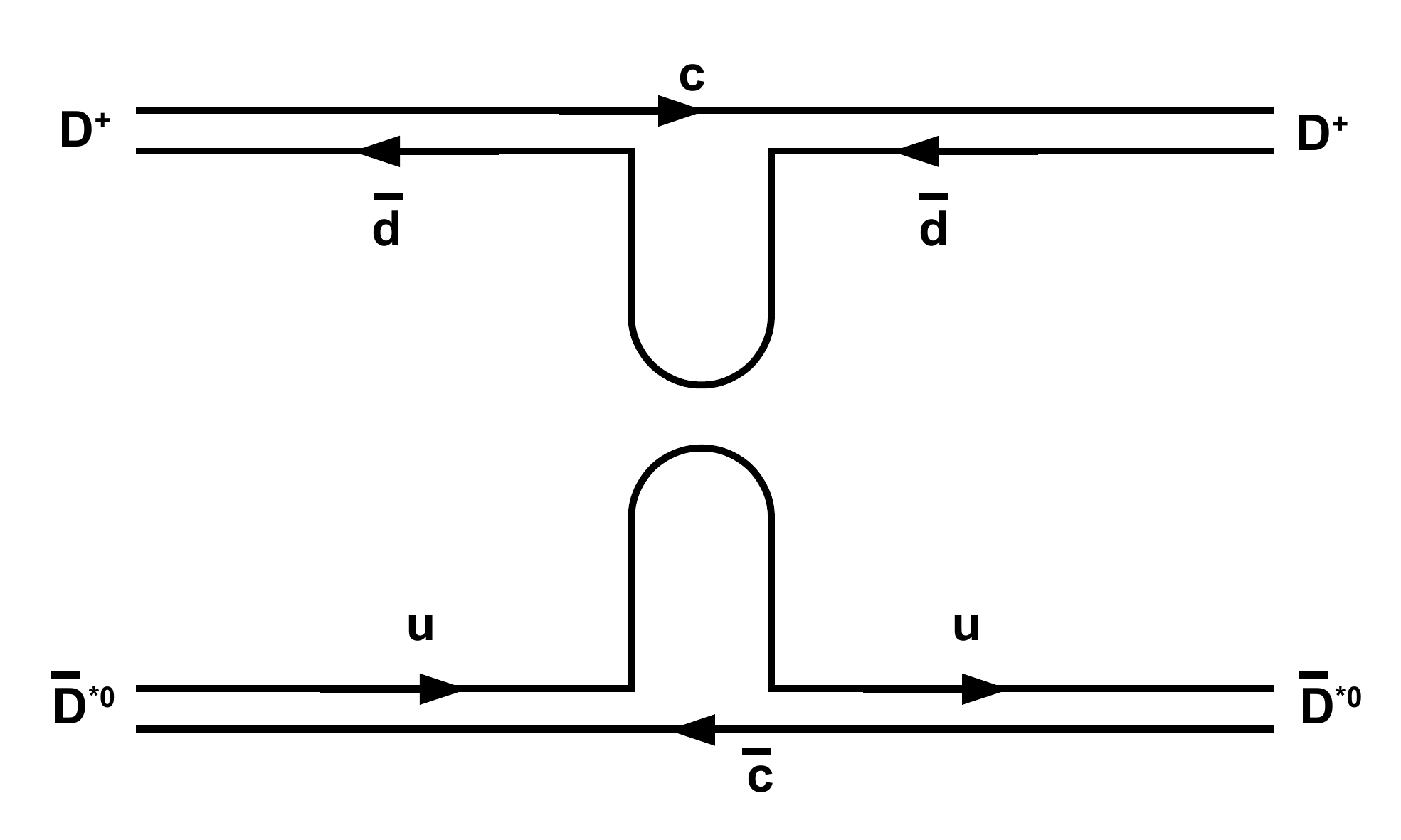}
\caption{Feynman diagram depicting the exchange of a light $q\bar{q}$ pair. A $d\bar{d}$ from the upper vertex is forced to convert into a $u\bar{u}$ pair in the lower one, evidencing an OZI forbidden mechanism.}
\label{fig:OZI}
\end{figure}

We want to study states of $I=1$ eventually generated by the $D\bar{D}^*$ interaction. To do this, we follow the approach of Ref. \cite{melafran}, starting from the observation that, as shown in  Fig. \ref{fig:OZI}, the exchange of a light meson is OZI forbidden, since a $d\bar{d}$ state exchange is forced to be converted into a $u\bar{u }$ state. This means that the contributions coming from $\rho$ and $\omega$ exchange cancel when taking equal masses and the same happens in the case of $\pi$, $\eta$, $\eta'$ mesons  if equal masses are taken, or  for large momenta bigger than the mass of the mesons. 

Thus, we evaluate the heavy vector exchange, where the OZI restriction no longer holds.

%******************************************************************************************************************************************************************

\subsection{Vector exchange}

We want to study the interaction between pseudoscalar mesons and vectors in the charm sector. In particular, we are interested in possible states with quantum numbers $C=0$, $S=0$ and $I=1$.

 In this sector, it is possible to distinguish between positive and negative $G$-parity combinations. In the case of positive $G$-parity ($I^G(J^{PC})=1^+(1^{+-}))$, six possible channels can contribute: $\pi\omega$ $\eta\rho$, $(\bar{K}{K}^*+c.c.)/\sqrt{2}$, $(\bar{D}{D}^*+c.c.)/\sqrt{2}$, $\eta_c\rho$ and $\pi J/\Psi$ \cite{daniel}\footnote{Note we have $C\rho^0=-\rho^0$, $C\rho^+=-\rho^-$, $C\rho^-=-\rho^+$.}. However, we will only take into account the last three: since we are investigating the energy region around $3900$ MeV, the $\pi\omega$ and $\eta\rho$ channels, whose thresholds are at much smaller energies, will only slightly affect the results. For negative $G$-parity  ($I^G(J^{PC})=1^-(1^{++}))$, we will only account for the $(\bar{D}{D}^*-c.c.)/\sqrt{2}$ channel, since the $(\bar{K}{K}^*-c.c.)/\sqrt{2}$ and $\pi\rho$ are too far from the energy values we are interested in.

In order to study the $PV\rightarrow PV$ interaction, we use the hidden gauge symmetry \cite{hidden1, hidden2, hidden3, hidden4} extended to SU(4) \cite{roca}, which is a very useful tool when dealing with vector mesons. We need the Lagrangian describing the $VPP$ vertex, given by
\begin{equation}
\mathcal{L}_{VPP}=-ig\langle V^{\mu}[P,\partial_{\mu}P]\rangle\ ,
\label{eq:VVPlag}
\end{equation}
where the symbol $\langle ~ \rangle$ stands for the trace of SU(4). The matrix $P$ contains the 15-plet of the pseudoscalar mesons written in the physical basis in which $\eta$, $\eta'$ mixing is considered \cite{gamphi3770}, 
\begin{equation}
P=\left(
\begin{array}{cccc}
\frac{\eta}{\sqrt{3}}+\frac{\eta'}{\sqrt{6}}+\frac{\pi^0}{\sqrt{2}} & \pi^+ & K^+&\bar{D}^0\\
\pi^- &\frac{\eta}{\sqrt{3}}+\frac{\eta'}{\sqrt{6}}-\frac{\pi^0}{\sqrt{2}} & K^{0}&D^-\\
K^{-} & \bar{K}^{0} &-\frac{\eta}{\sqrt{3}}+\sqrt{\frac{2}{3}}\eta'&D^-_s\\
D^0&D^+&D^+_s&\eta_c
\end{array}
\right)\ ,
\label{eq:pfields}
\end{equation}
while $V_{\mu}$ is given by 
\begin{equation}
V_\mu=\left(
\begin{array}{cccc}
\frac{\omega}{\sqrt{2}}+\frac{\rho^0}{\sqrt{2}} & \rho^+ & K^{*+}&\bar{D}^{*0}\\
\rho^- &\frac{\omega}{\sqrt{2}}-\frac{\rho^0}{\sqrt{2}} & K^{*0}&D^{*-}\\
K^{*-} & \bar{K}^{*0} &\phi&D^{*-}_s\\
D^{*0}&D^{*+}&D^{*+}_s&J/\psi
\end{array}
\right)_\mu\ .
\label{eq:vfields}
\end{equation}
%Eq. \eqref{eq:vfields}
The coupling constant is $g=M_{V}/2f_{\pi}$, with $f_{\pi}=93$ MeV the pion decay constant and $M_V\simeq 800$ MeV.

For the three vector vertex,  we use the Lagrangian
\begin{equation}
\mathcal{L}_{VVV}=ig\langle (V^{\mu}\partial_{\nu}V_{\mu}-\partial_{\nu}V_{\mu}V^{\mu})V^{\nu}\rangle\ ,
\label{eq:VVVlag}
\end{equation}
where $V_{\mu\nu}$ is defined as
\begin{equation}
V_{\mu\nu}=\partial_{\mu}V_{\nu}-\partial_{\nu}V_{\mu}-ig[V_{\mu},V_{\nu}]\ .
\label{eq:vectensor}
\end{equation}

The Lagrangians in Eqs. \eqref{eq:VVPlag} and \eqref{eq:VVVlag} produce the $PV\rightarrow PV$ interaction by means of the exchange of one vector meson. The resulting amplitudes are identical to those obtained with the chiral lagrangian of \cite{wise}. In Refs. \cite{daniel,roca2} these amplitudes are explicitly evaluated and projected in $s$-wave, with the result
\begin{equation}
V_{ij}(s)=-\frac{\vec{\epsilon}\ \vec{\epsilon}\ '}{8f^2}\,\mathcal{C}_{ij}\left[3s-(M^2+m^2+M'^2+m'^2)- \frac{1}{s}(M^2-m^2)(M'^2-m'^2)\right]\ .
\label{eq:VVPPamp}
\end{equation}
The masses $M$ ($M'$) and $m$ ($m'$) in Eq. \eqref{eq:VVPPamp} correspond to the initial (final) vector meson and pseudoscalar meson, respectively, while the indices $i$ and $j$ represent the initial and final $VP$ channels.

In the case of positive $G$-parity, we will have a $3\times 3$ matrix for the coefficients $\mathcal{C}_{ij}$, 
\begin{equation}
\label{eq:cmatrix}
\mathcal{C}_{ij}=\left( \begin{array}{ccc}
-\psi\ \ \  & 2\sqrt{\frac{2}{3}}\gamma\ \ \  & 2\sqrt{\frac{2}{3}}\gamma \\
2\sqrt{\frac{2}{3}}\gamma\ \ \  & 0\ \ \  & 0 \\
2\sqrt{\frac{2}{3}}\gamma\ \ \  & 0\ \ \  & 0 \end{array} \right)\ ,
\end{equation}
with $\gamma=\left(\frac{m_L}{m_H}\right)^2$ and $\psi=-\frac{1}{3}+\frac{4}{3}\left(\frac{m_L}{m_H'}\right)^2$. The parameters $m_L$, $m_H$ and $m_H'$  are chosen of the order of magnitude of a light vector meson mass, of a charmed vector mass and of the $J/\psi$ mass. We take $m_L=800$, $m_H=2050$ MeV, and $m_H'=3000$ MeV as done in Ref. \cite{daniel}. The factors $\gamma$ and $\psi$ take into account the suppression due to the exchange of a heavy vector meson.
In the case of negative $G$-parity, only one channel is present, whose corresponding coefficient in Eq. \eqref{eq:VVPPamp} is $\mathcal{C}=-\psi$. In the language of vector meson exchange this means that a $J/\psi$ is exchanged. The potential of Eq. \eqref{eq:VVPPamp} comes from the expression $(p_1+p_1')(p_2+p_2')$, which is approximately $(p^{\ 0}_1+p'^{\ 0}_1)(p^{\ 0}_2+p'^{\ 0}_2)$. In \cite{xiaoliang} it was shown that this Weinberg-Tomozawa interaction should implement the factor $(p_1^0/m_{K^*})(p_2^0/m_{K^*})$ multiplying the $SU(3)$ value, that stems from the implementation of the heavy quark spin symmetry. The interaction used automatically incorporates this factor, so no changes are needed with respect to what was done in \cite{daniel}.

\begin{figure}
  \centering
  \subfigure[]{\label{fig:v11}\includegraphics[width=0.4965\textwidth]{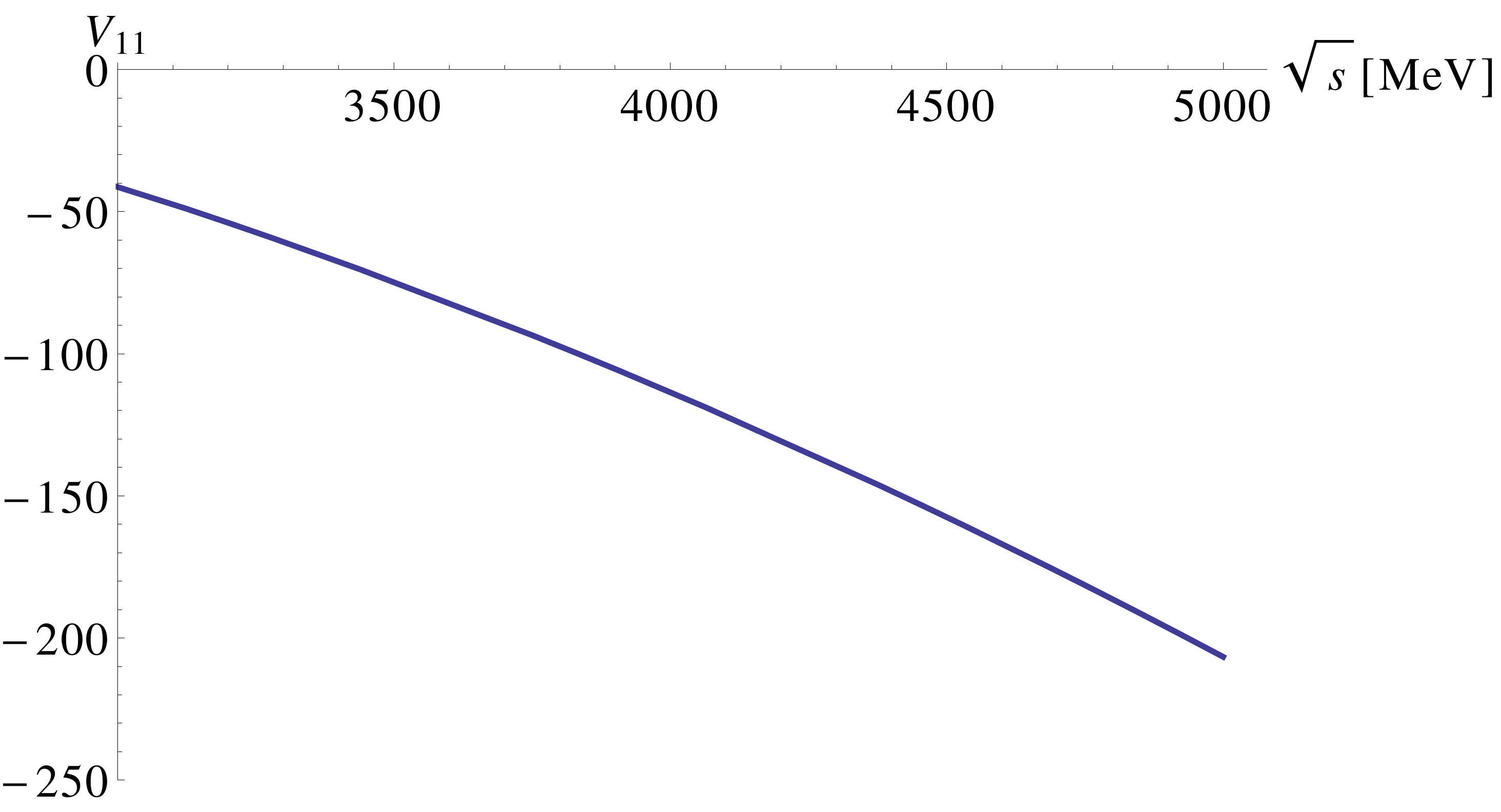}}                
   \subfigure[]{\label{fig:v12}\includegraphics[width=0.4965\textwidth]{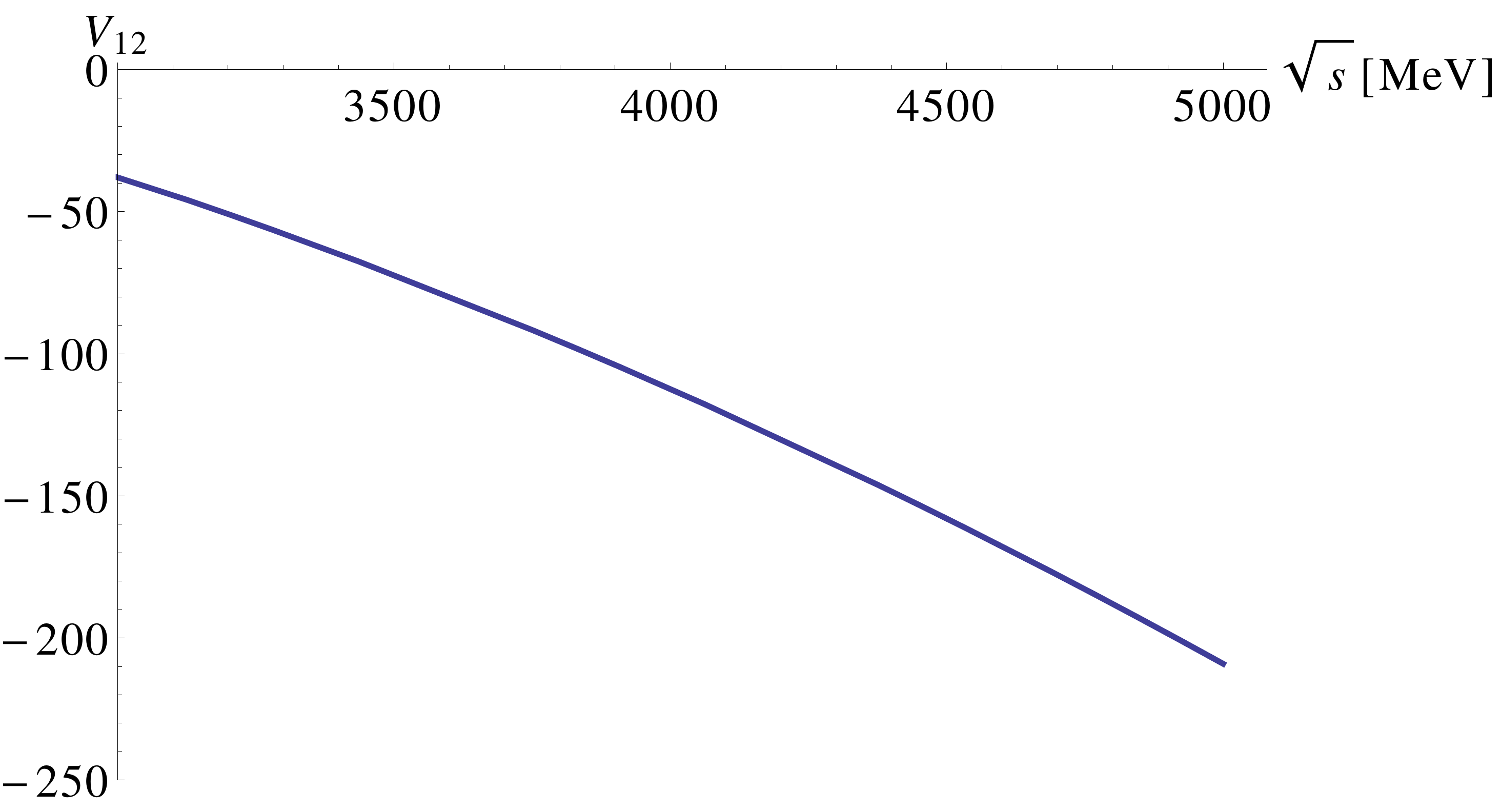}}
 \subfigure[]{\label{fig:v22}\includegraphics[width=0.4965\textwidth]{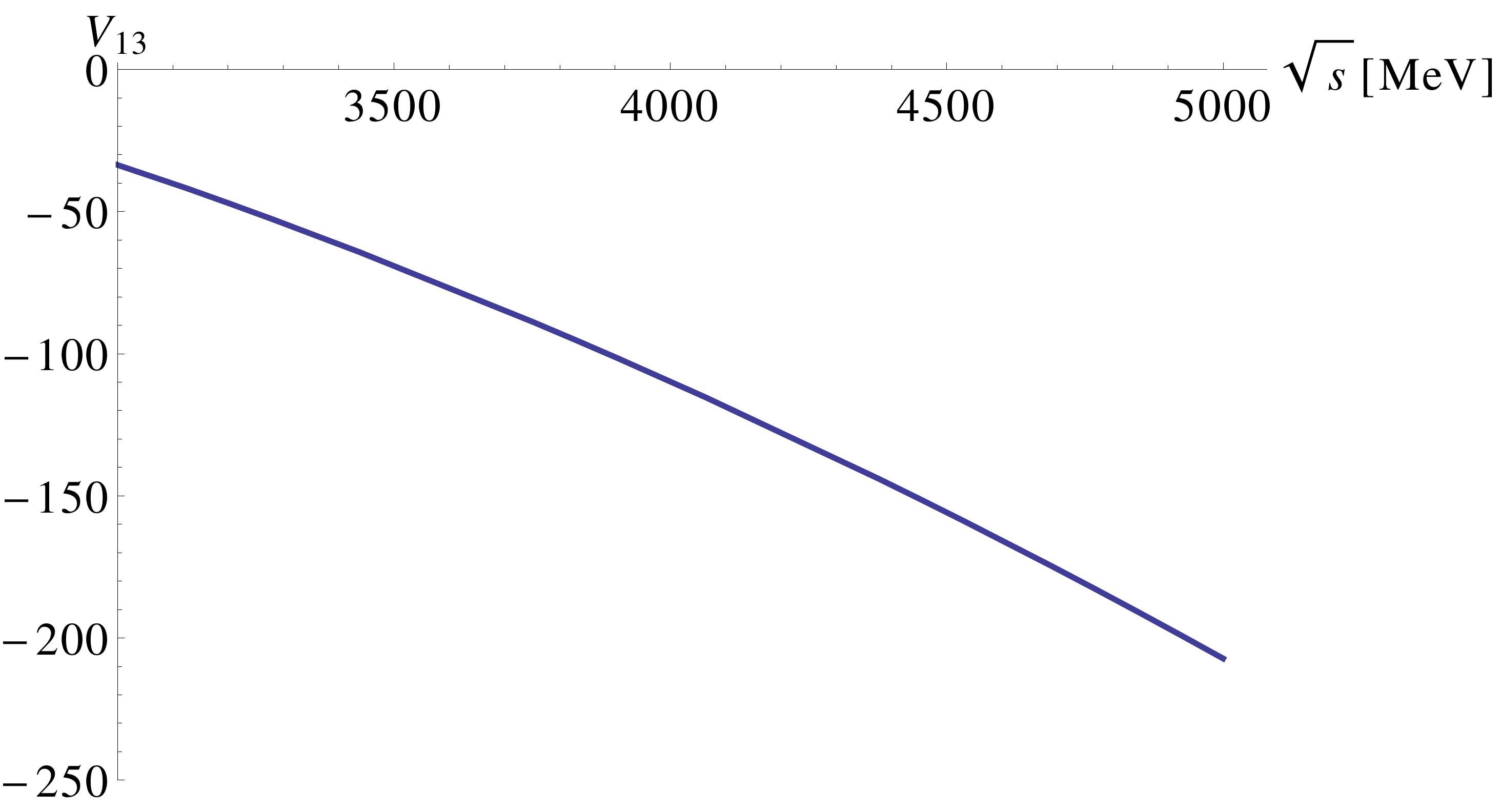}}
  \caption{Potentials $V_{D\bar{D}^*\rightarrow D\bar{D}^*}$  $(a)$, $V_{D\bar{D}^*\rightarrow\eta_C\rho}$ (b) and  $V_{D\bar{D}^*\rightarrow \pi J/\psi}$  $(c)$ as functions of the center of mass energy $\sqrt{s}$.}
\label{fig:PVPV}
\end{figure}

Eq. \eqref{eq:VVPPamp} provides the potential $V$ that must be used to solve the Bethe-Salpeter equation in coupled channels
\begin{equation}
T=(1-VG)^{-1}V\ ,
\label{eq:BS}
\end{equation}
removing the $\vec{\epsilon}\ \vec{\epsilon}\ '$ factor that factorizes also in $T$. The transition potentials $V_{ij}$ are shown in Fig. \ref{fig:PVPV}. 

The matrix $G$ is the diagonal loop function matrix whose elements are given by
\begin{equation}
G_l=i\int\frac{d^4q}{(2\pi)^4}\frac{1}{q^2-m^2+i\epsilon}\frac{1}{(q-P)^2-M_2^2+i\epsilon}\ ,
\label{eq:loopex}
\end{equation}
with  $m$ and $M$ the masses of the pseudoscalar and vector mesons, respectively, involved in the loop in the channel $l$ and $P$ the total four-momentum of the mesons.

 After the integration in $dq^0$,  Eq. \eqref{eq:loopex} becomes
\begin{equation}
\label{eq:1striemann}
G_l=\int\frac{d^3q}{(2\pi)^3}\,\frac{\omega_1+\omega_2}{2\omega_1\omega_2}\,\frac{1}{(P^0)^2-(\omega_1+\omega_2)^2+i\epsilon}\ ,
\end{equation}
with $\omega_1=\sqrt{m^2+\vec{q}^{\ 2}}$ and $\sqrt{M^2+\vec{q}^{\ 2}}$, which is regularized by means of a cutoff in the three-momentum $q_{max}$.

The function $G_l$ can be also written in dimensional regularization as
\begin{equation}
\begin{split}
\label{eq:loopexdm}
G_l&=\frac{1}{16\pi^2}(\alpha_l+\log\frac{m^2}{\mu^2}+\frac{M^2-m^2+s}{2s}\log\frac{M^2}{m^2}+\frac{p}{\sqrt{s}}(\log\frac{s-M^2+m^2+2p\sqrt{s}}{-s+M^2-m^2+2p\sqrt{s}}\\&+\log\frac{s+M^2-m^2+2p\sqrt{s}}{-s-M^2+m^2+2p\sqrt{s}}))\ ,
\end{split}
\end{equation}
where $p$ is the three-momentum of the mesons in the centre of mass
\begin{equation}
p=\frac{\sqrt{(s-(m+M)^2)(s-(m-M)^2)}}{2\sqrt{s}}\ .
\end{equation}
%********************************************************************************************************************************************************************

\subsection{One pseudoscalar exchange}
\label{onepseudoscalar}
\begin{figure}[htpb]
\centering
\includegraphics[scale=0.7]{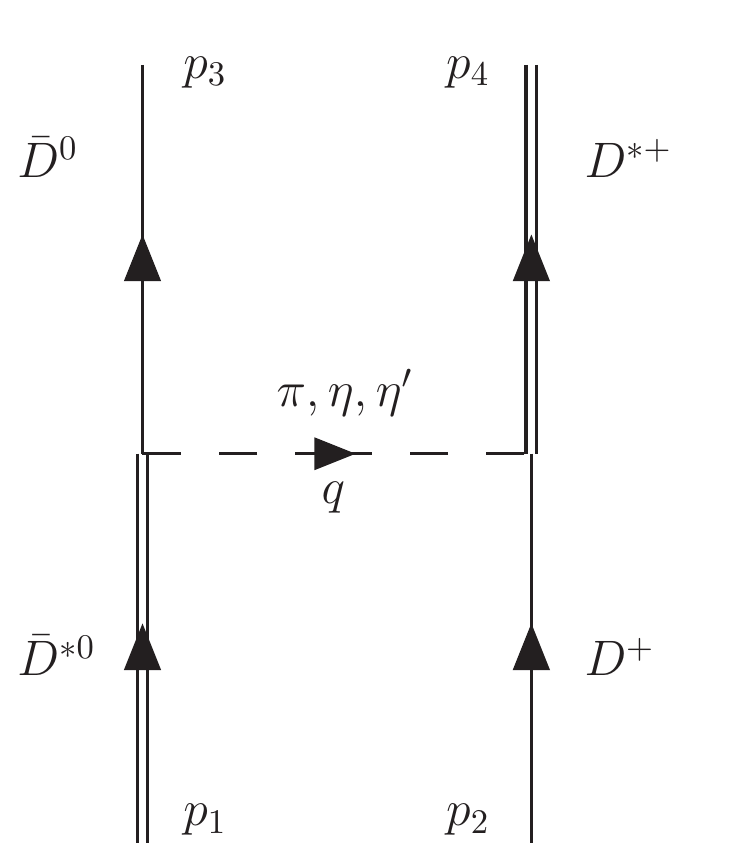}
\caption{Diagrammatic representation of the $D\bar{D}^*$ interaction via light pseudoscalar exchange.}
\label{fig:onemesondiag}
\end{figure}

In this section we proceed with the evaluation of the amplitude for the exchange of a single pseudoscalar ($\pi$, $\eta$, $\eta^{\prime}$). The process is depicted in Fig. \ref{fig:onemesondiag} and the Lagrangian we need to evaluate its amplitude is given by
\begin{equation}
\label{eq:lagrangianhg}
\mathcal{L}_{PPV}=-ig\ \langle V^\mu [P,\partial_\mu P]\rangle\ ,
\end{equation}
where the matrices $P$ and $V$ are given by Eqs. \eqref{eq:pfields} and \eqref{eq:vfields}. The constant $g$ is the strong coupling of the $D^*$ meson to $D\pi$, which in $SU(3)$ is equal to $4.16$. However this is in contradiction with the empirical value of $g\simeq 9$ needed to get the $D^{*}\rightarrow D\pi$ width. This apparent contradiction is settled in \cite{melafran} by looking at the $D^*\rightarrow D\pi$ decay using the impulse approximation at the quark level, assuming the heavy quarks as spectators. The standard normalization used for the meson fields at the macroscopic level (mesons, not quarks) demands that the $g\ \vec{\epsilon}\cdot\vec{q}$ operator that one has for the $D^{*0}$ decay at rest is normalized by an extra $m_{D^*}/m_{K^*}$ factor. This gives an effective $g$ constant for $D$, $D^*$ mesons of $9.40$. With this coupling we get a width of $71$ KeV for the $D^{*+}\rightarrow D^0\pi^+$ decay, which is in agreement with the more recent result of $(65\pm15)$ KeV of \cite{cleo}.

The state that we have is
\begin{equation}
\psi_1=\frac{1}{\sqrt{2}}\Big( D^{+}\bar{D}^{*0} +\bar{D}^{0}D^{*+}\Big) \ .
\label{eq:psi1}
\end{equation}
We can see that the pseudoscalar meson exchange with the interaction of Eq. \eqref{eq:lagrangianhg} mixes the first component of Eq. \eqref{eq:psi1} for the initial state with the second component of the same equation for the final state and vice-versa.

\begin{figure}[htpb]
\centering
\includegraphics[scale=0.35]{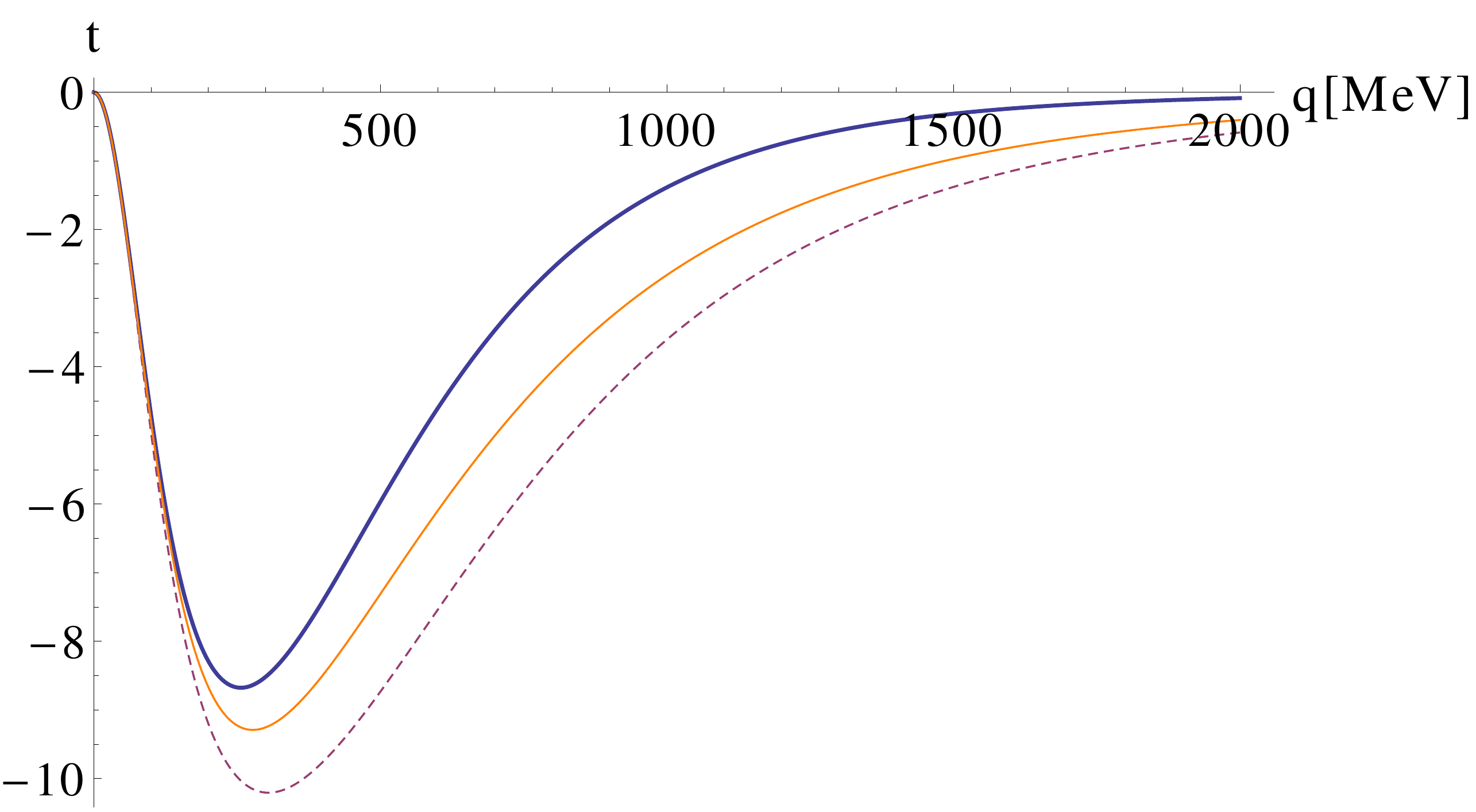}
\caption{One pseudoscalar exchange potential for the exchange of one pion (dashed line), $\pi$ plus $\eta$ (thin line) and $\pi$ plus $\eta$ plus $\eta^{\prime}$ (thick line) as functions of the transferred momentum $q$.}
\label{fig:onemeson}
\end{figure}

Altogether, using the P matrix of \cite{gamphi3770}, Eq. \eqref{eq:pfields}, that mixes $\eta$ and $\eta^{\prime}$, we find 
\begin{equation}
\begin{split}
t_{\psi_1 \rightarrow \psi_1}&=t_{\bar{D}^{*0}D^+\rightarrow \bar{D}^0 D^{*+}}=-4g^{2}\Big(-\frac{1}{2}\frac{1}{q^{2}-m_{\pi}^{2}+i\epsilon} + \frac{1}{6}\frac{1}{q^{2}-m_{\eta}^{2}+i\epsilon} +\frac{1}{3}\frac{1}{q^{2}-
m_{\eta^{\prime}}^{2}+i\epsilon}\Big)\\& \times(\epsilon_1 \cdot p_{3})(\epsilon_4 \cdot p_{2}) F^{2}(\vec{q\,})\, ,
\end{split}
\label{eq:onepseudo}
\end{equation}
where $m_{\pi}$ is the mass of the pion, $\epsilon_1$ and $\epsilon_4$ are the polarization vectors for the $\bar{D}^{*0}$ and $D^{*+}$ vector mesons, respectively and $F(\vec{q\,})$ is a form factor of the type $F(\vec{q\,})=\frac{\Lambda^2}{\Lambda^2+\vec{q\,}^{2}}$, with $\Lambda=1$ GeV, which is also used later in the two pion exchange. Considering that the masses are heavier than the external momenta, this implies the following approximations: $\epsilon_1 \cdot p_3=-\vec{\epsilon_1}\cdot \vec{p_3}$ and $\epsilon_4 \cdot p_2=-\vec{\epsilon_4}\cdot \vec{p_2}$. We use the Breit frame where
\begin{equation}
\begin{split}
&p_1\equiv (p_1^0,\vec{q}/2) \ ,\\
&p_2\equiv (p_2^0,-\vec{q}/2) \ ,\\
&p_3\equiv (p_3^0,-\vec{q}/2) \ ,\\
&p_4\equiv (p_4^0,\vec{q}/2) \ .
\end{split}
\end{equation}

Since we are doing an estimate, we have chosen $q^0\equiv 0$. We are dealing with $s$-waves, and this allows us to use $q_{i}q_{j}\rightarrow \frac{1}{3} \vec{q\,}^2 \delta_{ij}$ and then to rewrite the amplitude of Eq. \eqref{eq:onepseudo} due to the exchange of a pseudoscalar meson ($\pi$, $\eta$, and $\eta^{\prime}$) as
\begin{equation}
t_{\bar{D}^{*0}D^+\rightarrow \bar{D}^0 D^{*+}}=\frac{g^{2}}{3}\vec{q\,}^2\Big(-\frac{1}{2}\frac{1}{\vec{q\,}^2+m_{\pi}^{2}+i\epsilon} + \frac{1}{6}\frac{1}{\vec{q\,}^2+m_{\eta}^{2}+i\epsilon} +\frac{1}{3}\frac{1}{\vec{q\,}^2+m_{\eta^{\prime}}^{2}+i\epsilon}\Big)F^{2}(\vec{q\,}) \ .
\label{eq:onemeson}
\end{equation}

In Fig. \ref{fig:onemeson} we show the contributions coming from Eq. \eqref{eq:onemeson} for the exchange of one pion (dashed line), $\pi$ plus $\eta$ (thin line) and $\pi$ plus $\eta$ plus $\eta^{\prime}$ (thick line) as functions of the transferred momentum $q$. We can see a partial cancellation between the three contributions, which becomes very effective at large momenta. 

\begin{figure}[htpb]
\centering
\includegraphics[scale=0.35]{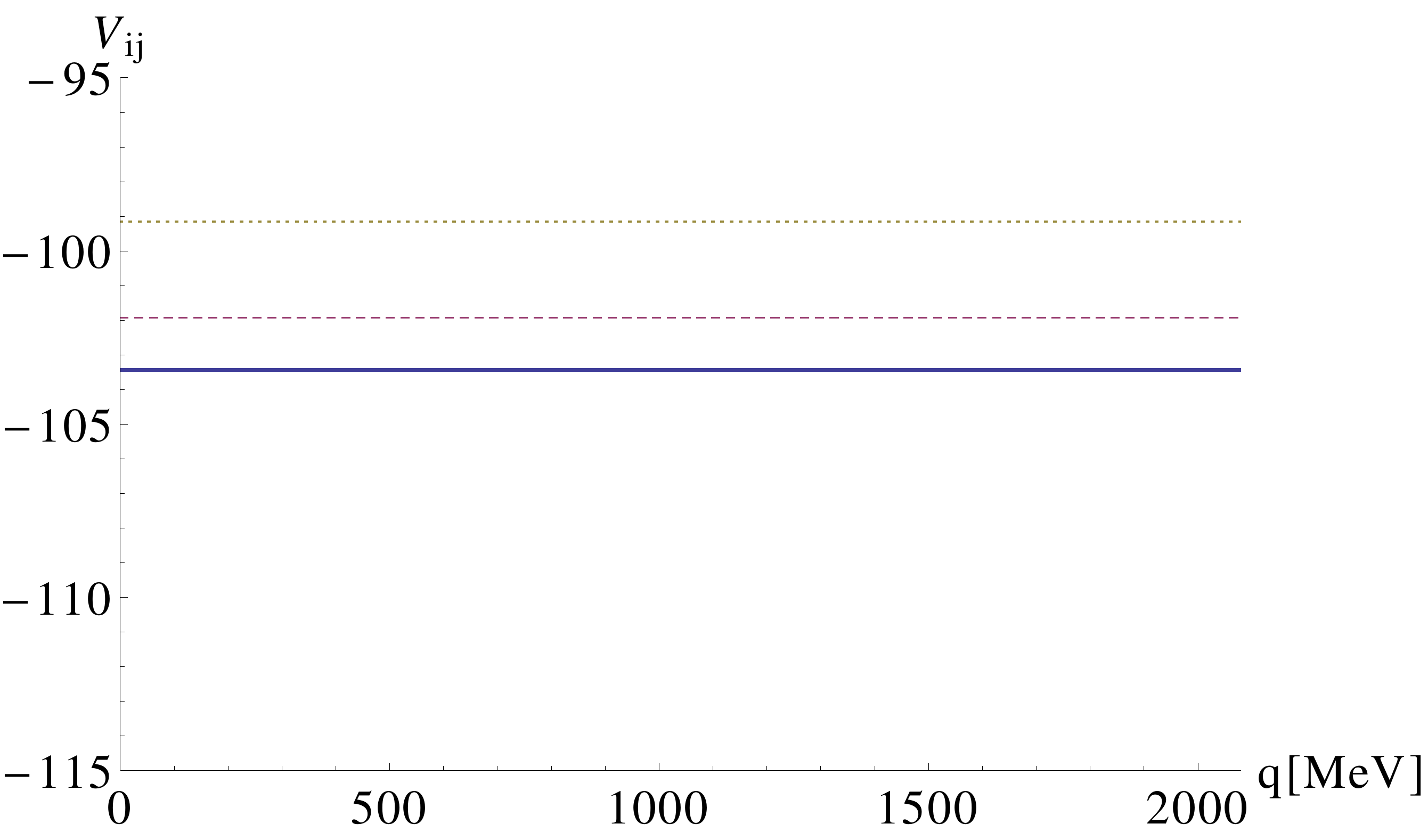}
\caption{Vector exchange potentials $V_{11}$ (thick line), $V_{12}$ (dashed line) and $V_{13}$ (dotted line) as functions of the transferred momentum $q$.}
\label{fig:vectorq}
\end{figure}

It is interesting to compare the contribution of Fig. \ref{fig:onemeson} with the one due to vector exchange which we plot in Fig. \ref{fig:vectorq}. Recall that the use of the vector exchange potential in $V$ of Eq. \eqref{eq:BS}, together with a $G$ function regularized with a cutoff $q_{max}$, is equivalent to using a potential $V(\vec{p},\vec{p\,}^{\prime})=V\theta(q_{max}-\vec{p\,})\theta(q_{max}-\vec{p\,}^{\prime})$ \cite{Gamermann:2009uq}. Assuming $\vec{p\,} \simeq 0$, then $\vec{p\,}^{\prime}$ takes the place of $\vec{q\,}$, and this allows a proper comparison, recalling that $q_{max}$, to be used later, is of the order of $770$ MeV. We can safely conclude that the exchange of pseudoscalar mesons is very small compared to the vector exchange.

%**************************************************************************************************
\subsection{The $D\bar{D}^*$ interaction by means of $\sigma$ exchange}

In  Ref. \cite{toki}, the exchange of two correlated (interacting) pions in  the $NN$ interaction was studied. In Ref. \cite{melafran}, the same idea was extended to the case of $D^*\bar{D}^*$. We apply the same formalism here to study the $D\bar{D}^*$ interaction.

\begin{figure}[htpb]
\centering
\includegraphics[scale=0.56]{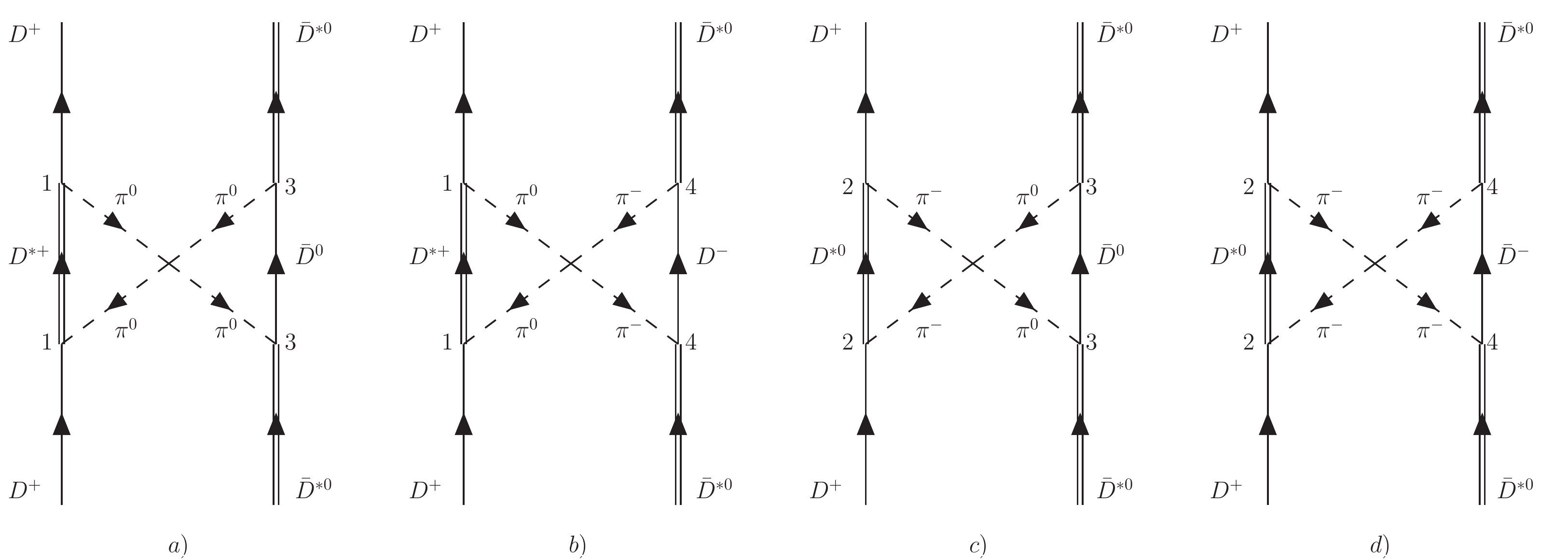}
\caption{Lowest order $\pi\pi$ interaction in the $I=1$ channel for $D\bar{D}^*\rightarrow D\bar{D}^*$.}
\label{fig:sigmadiag}
\end{figure}

The diagrams contributing to this process are shown in Fig. \ref{fig:sigmadiag}. Each one of  them contains four $PPV$ vertices easily evaluated by means of the local hidden gauge. The crossing of the pion lines indicates that we have there the $\pi\pi$ scattering amplitude that contains the $\sigma$ pole ($f_0(500)$). In addition to the $PPV$ vertex we could also consider the $PVV$ one, allowing then two $D^*$ intermediate states, but the anomalous character of the $PVV$ vertex renders these terms smaller than those considered here.
The Lagrangian we need to evaluate the amplitudes is given by Eq. \eqref{eq:lagrangianhg}.

As found in Ref. \cite{melafran}, the amplitude for the diagrams in Fig. \ref{fig:sigmadiag} can be written as
\begin{equation}
-it_{\sigma}=-i\ V_A\, V_B\ \frac{3}{2}\ t_{\pi\pi\rightarrow\pi\pi}^{I=0}\ ,
\label{eq:ampl_sigma}
\end{equation}
where
\begin{equation}
t_{\pi\pi\rightarrow\pi\pi}^{I=0}=-\frac{1}{f^2}\ \frac{s-\frac{m_{\pi}^2}{2}}{1+\frac{1}{f^2}(s-\frac{m_{\pi}^2}{2})G(s)}\ 
\label{eq:bs-ampl}
\end{equation}
is the on-shell part of the isoscalar amplitude for the $\pi\pi$ interaction summed up to all orders in the unitary approach \cite{oller}. The function $G(s)$ in Eq. \eqref{eq:ampl_sigma} is the two pion loop function, conveniently regularized \cite{toki},
\begin{equation}
G(s)=i\int \frac{d^4q}{(2\pi)^4}\,\frac{1}{q^2-m_{\pi}^2+i\epsilon}\,\frac{1}{(P-q)^2-m_{\pi}^2+i\epsilon}\ ,
\label{eq:loopf}
\end{equation}
with $P$ the total momentum of the two pion system, $P^2=s$.
\begin{figure}[htpb]
\centering
\includegraphics[scale=0.5]{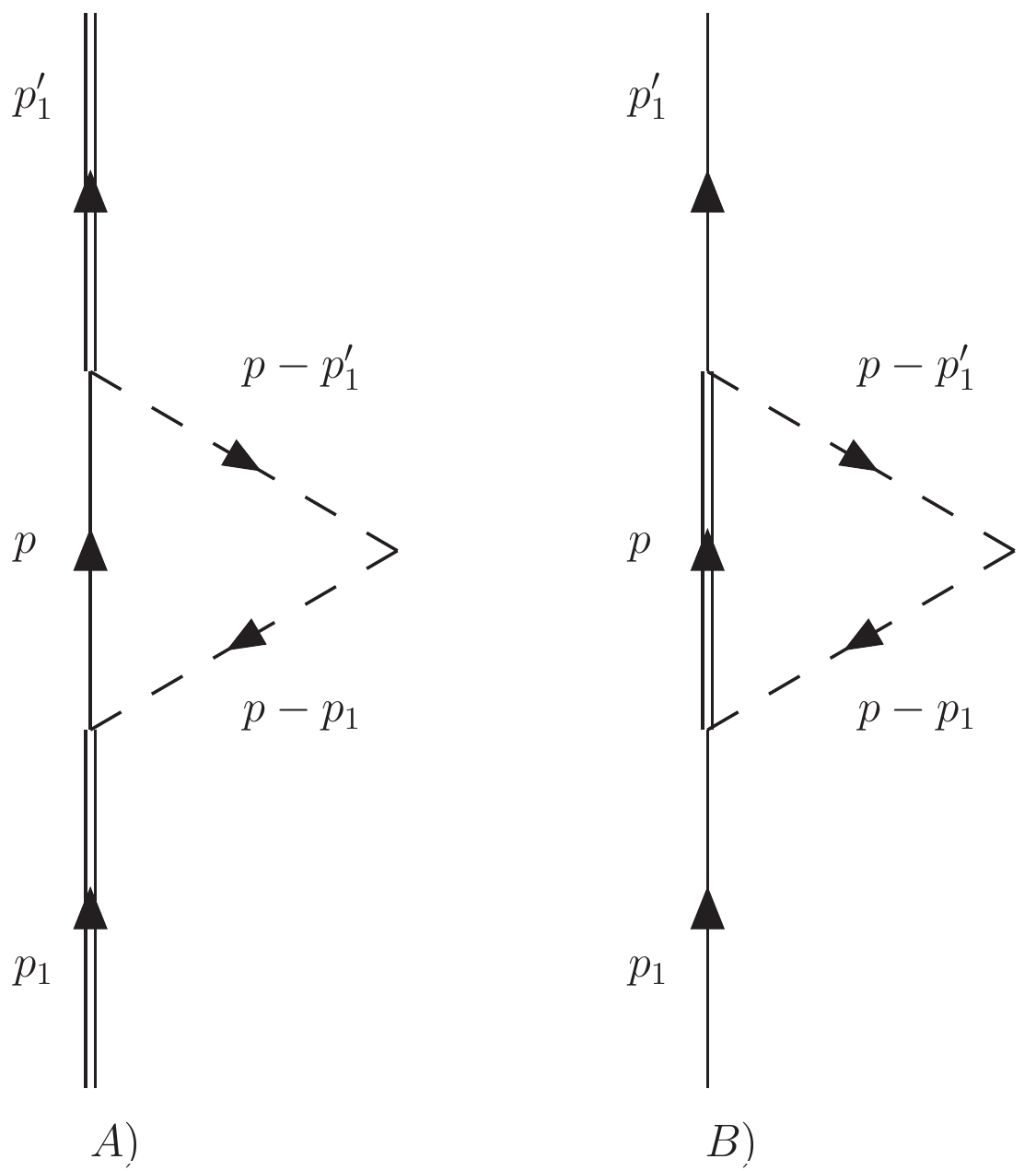}
\caption{Two pion exchange triangle vertices, $V_A$ in Fig. A) and $V_B$ in Fig. B).}
\label{fig:V}
\end{figure}

The two factors $V_A$ and $V_B$ in Eq. \eqref{eq:ampl_sigma} represent the contributions coming from the two triangular loops in the diagrams, which are shown in Fig. \ref{fig:V}. The detailed derivation for $V_A$ can be found in Ref. \cite{melafran}. We use again the Breit reference frame in which
\begin{equation}
\begin{split}
&p_1\equiv(p_1^0, \vec{q}/2)\ ,\\
&p_1'\equiv(p_1'^{\ 0}, -\vec{q}/2)\ ,\\
&p\equiv(p^0, \vec{p}\,)\ ,
\label{eq:breit}
\end{split}
\end{equation}
where $\vec{q}$ is the three-momentum transferred in the process. Since there is no energy exchange, $s=-\vec{q}^{\ 2}$ in Eq. \eqref{eq:bs-ampl}. 

We can write 
\begin{equation}
\begin{split}
V_A&=ig^2\int\frac{d^4p}{(2\pi)^4}\epsilon_{\mu}(2p-p_1)^{\mu}\epsilon'_{\nu}(2p-p_1')^{\nu}\frac {1}{p^2-m^2_D+i\epsilon}\\&\times\frac{1}{(p-p_1)^2-m^2_{\pi}+i\epsilon}\,\frac{F}{(p-p_1')^2-m^2_{\pi}+i\epsilon}\ 
\label{eq:VA}
\end{split}
\end{equation}
and 
\begin{equation}
\begin{split}
V_B&=ig^2\int\frac{d^4p}{(2\pi)^4}\epsilon_{\mu}(p-2p_1)^{\mu}\epsilon_{\nu}(p-2p_1')^{\nu}\frac {1}{p^2-m^2_{D^*}+i\epsilon}\\&\times\frac{1}{(p-p_1)^2-m^2_{\pi}+i\epsilon}\,\frac{F}{(p-p_1')^2-m^2_{\pi}+i\epsilon}\ ,
\label{eq:VB}
\end{split}
\end{equation}
with $m_D$ and $m_{D^*}$ the masses of the $D$ and $D^*$ mesons respectively.  The factor $F$ in both equations is the product of two static form factors 
\begin{equation}
F=F_1(\vec{p}+\frac{\vec{q}}{2})\, F_2(\vec{p}-\frac{\vec{q}}{2})=\frac{\Lambda^2}{\Lambda^2+(\vec{p}+\frac{\vec{q}}{2})^2}\,\frac{\Lambda^2}{\Lambda^2+(\vec{p}-\frac{\vec{q}}{2})^2}\ ,
\label{eq:ff}
\end{equation}
with $\Lambda=1$ GeV, and, together with a cutoff in the space of intermediate states ($p_{max}=2$ GeV), it is needed to regularize the integrals in Eqs. \eqref{eq:VA} and \eqref{eq:VB} which are logarithmically divergent. This was the cutoff needed in Ref. \cite{toki} to obtain the result of the empirical $\sigma$ exchange at large distances.

In Ref. \cite{melafran} it was found that, using the Lorentz conditions  $\epsilon_{\mu}\,p_1^{\mu}=0$ and $\epsilon'_{\nu}\,p_1'^{\nu}=0$, the final expression for $V_A$ has the form
\begin{equation}
V_A=\epsilon_{\mu}\epsilon'_{\nu}(ag^{\mu\nu}+cp_1'^{\mu}p_1^{\nu})\ ,
\label{eq:VA2}
\end{equation}
where
\begin{equation}
\begin{split}
&a=\frac{-Ym^2_{D^*}+Z(p_1p_1')+X(m^4_{D^*}-(p_1p_1')^2)}{2(m^4_{D^*}-(p_1p_1')^2)}\ ,\\
&c=\frac{-3Ym^2_{D^*}(p_1p_1')+X(p_1p_1')(m^4_{D^*}-(p_1p_1')^2)+Z(m^4_{D^*}+2(p_1p_1')^2)}{2(m^4_{D^*}-(p_1p_1')^2)^2}\ ,
\label{eq:ac}
\end{split}
\end{equation}
and 
\begin{equation}
\begin{split}
&X=4g^2I_1+4g^2m_D^2I_2\ ,\\
&Y=8g^2p_1^{0\,2}I_1+8g^2I_3\ ,\\
&Z=8g^2p_1^{0\,2}I_1+8g^2I_4\ .
\label{eq:ac2}
\end{split}
\end{equation}

For low three momenta of the external vector mesons compared to their masses, which is assumed here, where $\epsilon^0\equiv0$, and also low momenta of the external $D$, Eq. \eqref{eq:VA2} gives $V_A=-a \vec{\epsilon}\ \vec{\epsilon}\ '$, and the factor $\vec{\epsilon}\ \vec{\epsilon}\ '$ factorizes in the amplitude $t_{\sigma}$.

The four integrals in the equations above, $I_1$, $I_2$, $I_3$ and $I_4$, after performing the integration in $dp^0$, which can be done analytically using Cauchy's theorem, have the following expressions:
\begin{equation}
\begin{split}
&I_1=\int\frac{d^3p}{(2\pi)^3}\,\frac{\omega_1+\omega_2}{2\omega_1\omega_2}\,\frac{1}{-\vec{q}^{\,2}-(\omega_1+\omega_2)^2}\,F\ ,\\
&I_2=\int\frac{d^3p}{(2\pi)^3}\,\frac{1}{2E_D}\,\frac{1}{2\omega_1}\,\frac{1}{\omega_2}\,\frac{1}{\omega_1+\omega_2}\,\frac{\omega_1+\omega_2+E_D-m_{D^*}}{E_D+\omega_1-m_{D^*}-i\epsilon}\,\frac{1}{E_D+\omega_2-m_{D^*}-i\epsilon}\,F\ ,\\
&I_3=\int\frac{d^3p}{(2\pi)^3}\,\frac{1}{2E_D}\,\frac{1}{2\omega_1}\,\frac{1}{\omega_2}\,\frac{1}{\omega_1+\omega_2}\,\frac{\omega_1+\omega_2+E_D-m_{D^*}}{E_D+\omega_1-m_{D^*}-i\epsilon}\,\frac{(\vec{p}^{\,2}+m_D^2)p_1^{0\,2}+(\vec{p}\,\frac{\vec{q}}{2})^2}{E_D+\omega_2-m_{D^*}-i\epsilon}\,F\ ,\\
&I_4=\int\frac{d^3p}{(2\pi)^3}\,\frac{1}{2E_D}\,\frac{1}{2\omega_1}\,\frac{1}{\omega_2}\,\frac{1}{\omega_1+\omega_2}\,\frac{\omega_1+\omega_2+E_D-m_{D^*}}{E_D+\omega_1-m_{D^*}-i\epsilon}\,\frac{(\vec{p}^{\,2}+m_D^2)p_1^{0\,2}-(\vec{p}\,\frac{\vec{q}}{2})^2}{E_D+\omega_2-m_{D^*}-i\epsilon}\,F\ ,
\label{eq:I2}
\end{split}
\end{equation}
where $\omega_1=\sqrt{(\vec{p}+\vec{q}/2)^2+m_{\pi}^2}$, $\omega_2=\sqrt{(\vec{p}-\vec{q}/2)^2+m_{\pi}^2}$ and $E_D=\sqrt{\vec{p}^{\ 2}+m_D^2}$ are the energies of the two pions and of the $D$ meson involved in the loop, respectively, and $m_{D^*}$ is the mass of the $\bar{D}^*$ meson. The former equations are obtained taking only the positive energy part of the $D$ propagator $[(p^0-E_D)2E_D]^{-1}$, which is a very good approximation given the large mass of the $D$.

In the case of $V_B$, after some simple algebra, we obtain 
\begin{equation}
\begin{split}
V_B&=g^2I_1+g^2\left[2(m_D^2-m_{\pi}^2)-4p_1p_1'-\frac{(m_D^2-m_{\pi}^2)^2}{m_{D^*}^2}+m_{D^*}^{2}\right]I_5\\&-2g^2\left[1+\frac{m_D^2-m_{\pi}^2}{m_{D^*}^2}\right]I_6+g^2\frac{1}{m_{D^*}^2}I_7\ ,
\label{eqVB2}
\end{split}
\end{equation}
where
\begin{equation}
\begin{split}
&I_5=\int\frac{d^3p}{(2\pi)^3}\,\frac{1}{2E_V}\,\frac{1}{2\omega_1}\frac{1}{\omega_2}\,\frac{1}{\omega_1+\omega_2}\,\frac{\omega_1+\omega_2+E_V-m_D}{E_V+\omega_1-m_D}\,\frac{F}{E_V+\omega_2-m_D}\ ,\\
&I_6=\int\frac{d^3p}{(2\pi)^3}\,\frac{1}{2E_V}\,\frac{F}{\omega_1}\,\frac{\omega_1+E_V}{p_1^{0\,2}-(\omega_1+E_V)^2}\ ,\\
&I_7=\int\frac{d^3p}{(2\pi)^3}\,\frac{F}{2E_V}\ ,
\label{eq:IB2}
\end{split}
\end{equation}
where $E_V=\sqrt{\vec{p}^{\ 2}+m_{D^*}^2}$. Once again the non relativistic propagator for the  intermediate $D^*$ has been taken to get the former equations.

\begin{figure}[htpb]
\centering
\includegraphics[scale=0.4]{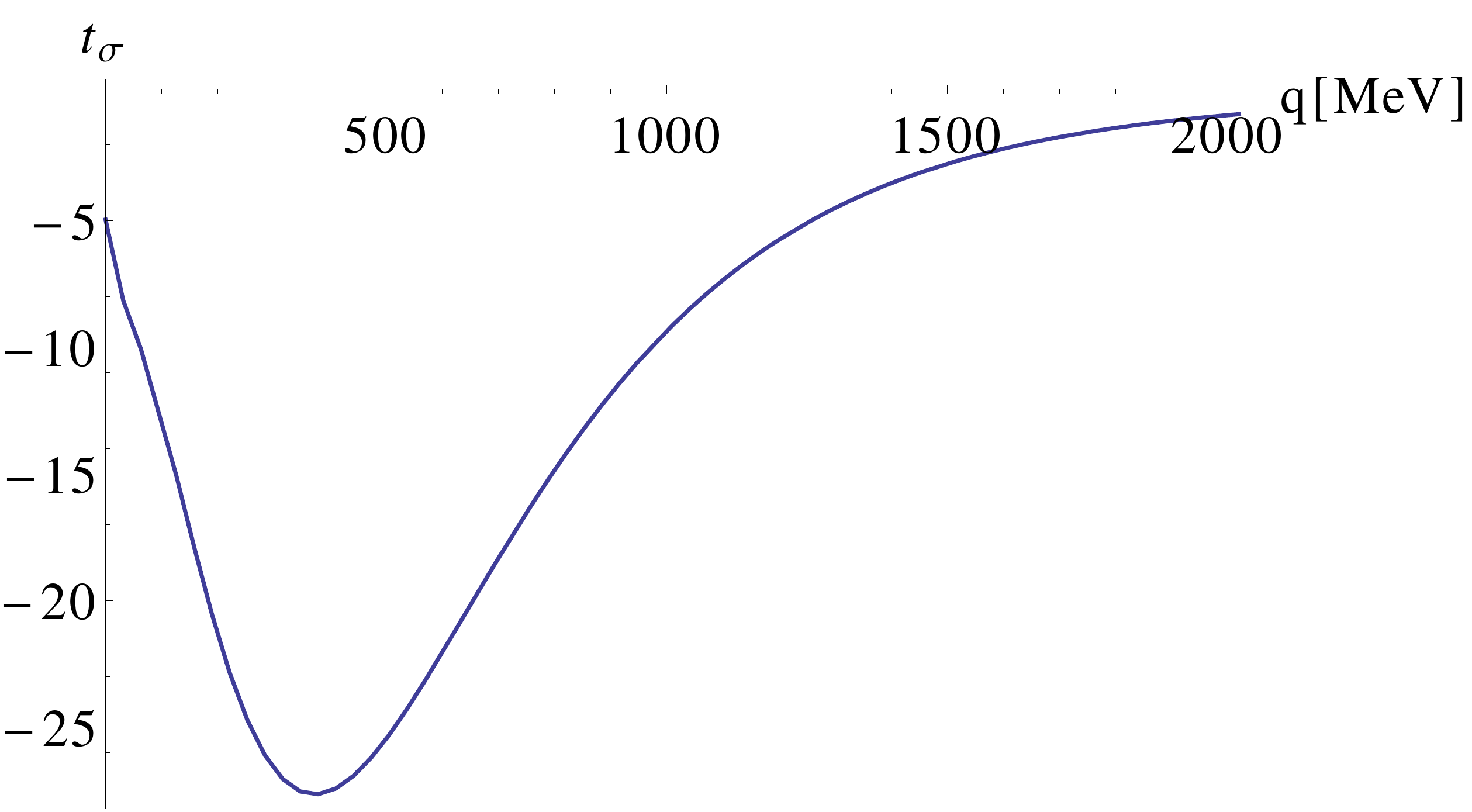}
\caption{Potential $t_{\sigma}$ as a function of the momentum transferred in the process.}
\label{fig:tsigma}
\end{figure}

The potential $t_{\sigma}$ of Eq. \eqref{eq:ampl_sigma} as a function of the transferred momentum $\vec{q}$ is plotted in Fig. \ref{fig:tsigma}. One can observe, from the comparison with Fig. \ref{fig:vectorq}, that this contribution is reasonably smaller than that of vector exchange, but the faster fall as a function of $\vec{q\,}$ makes its contribution less relevant, as we shall discuss later.

%*************************************************************************************************
\subsection{Uncorrelated crossed two pion exchange}
Now we study the case of the exchange of two non interacting pions. Only the two crossed diagrams $a)$ and $d)$ of Fig. \ref{fig:sigmadiag} contribute to the process. 

\begin{figure}[htpb]
\centering
\includegraphics[scale=0.5]{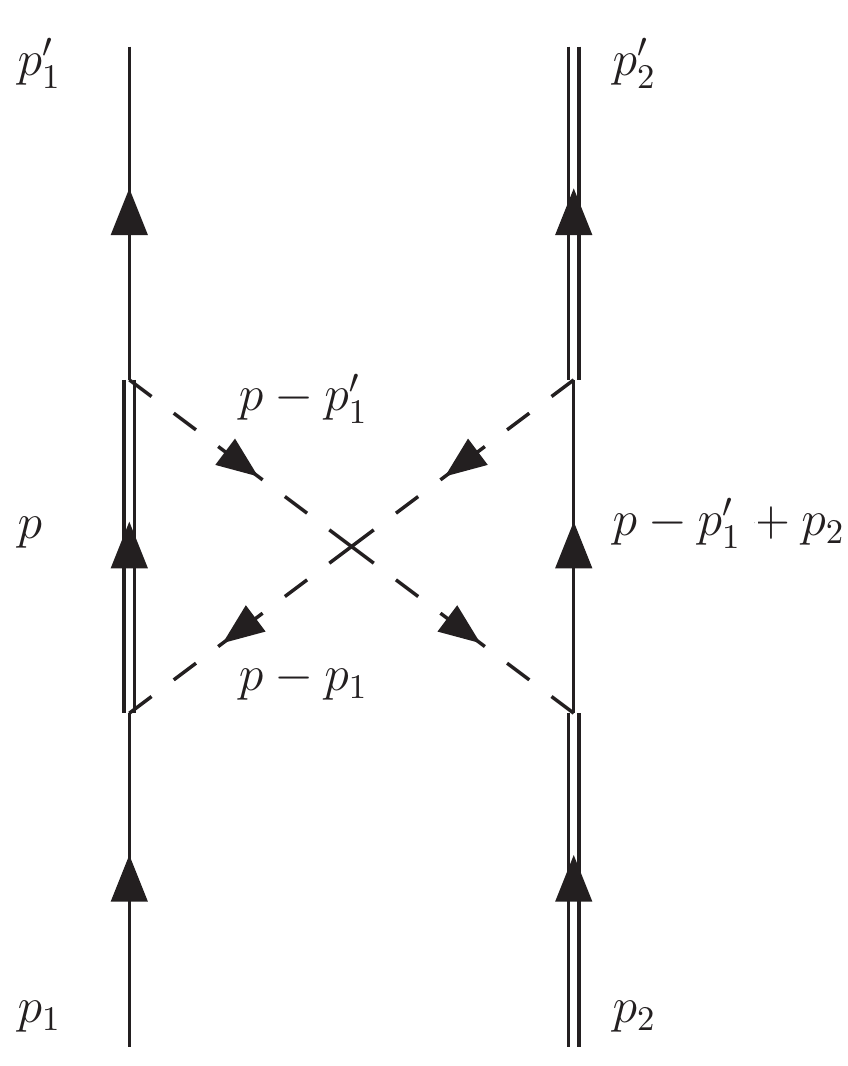}
\caption{Momenta assignment in the two pion exchange in $D\bar{D}^*\rightarrow D\bar{D}^*$.}
\label{fig:momenta}
\end{figure}

The evaluation of the amplitude is completely analogous to the case of the $D^*\bar{D^*}$ interaction evaluated in Ref. \cite{melafran}, but recalling that now we have one propagator for $D$ meson and one for the $D^*$. We obtain, with the momenta assignment of Fig. \ref{fig:momenta},
\begin{equation}
\begin{split}
t&=\frac{5}{4}ig^4\int\frac{d^4p}{(2\pi)^4}\,\epsilon_{\mu}(2p_1-p)^{\mu}\epsilon_{\nu}(2p_1'-p)^{\nu}\epsilon_{\alpha}'(2p-2p_1'+p_2)^{\alpha}\epsilon_{\beta}''(2p-p_1'-p_1+p_2)^{\beta}\\&\times\frac{ F^2}{p^2-m^2_{D^*}+i\epsilon}\,\frac{1}{(p-p_1'+p_2)^2-m_D^2+i\epsilon}\,\frac{1}{(p-p_1)^2-m^2_{\pi}+i\epsilon}\,\frac{1}{(p-p_1')^2-m^2_{\pi}+i\epsilon}\ ,
\label{eq:ampl_rel}
\end{split}
\end{equation}
where $\epsilon$ is the polarization four-vector corresponding to the vector meson in the triangular loop, while $\epsilon'$ and $\epsilon''$ correspond to the vector mesons in the external legs of the diagram.

Once again, we take the positive energy part of the $D$ and $D^*$ propagators and for the external vectors we assume small three-momenta, hence $\epsilon^0\equiv 0$. We also assume that $4\vec{p}^{\,2}\gg\vec{q}^{\,2}/4$. Thus, applying the completeness condition for the polarization vector, we can rewrite Eq. \eqref{eq:ampl_rel} as
\begin{equation}
\begin{split}
t&=\frac{5}{4}ig^4\,\frac{1}{2}\,\vec{\epsilon}\ '\ \vec{\epsilon}\ ''\int\frac{d^4p}{(2\pi)^4}\,(\vec{p}^{\ 2}-\vec{q}^{\ 2})\left[(4\vec{p}^{\ 2}-\frac{\vec{q}^{\ 2}}{4})-\frac{1}{\vec{q}^{\ 2}}\left[(2\vec{p}\,\vec{q}\,)^2-\frac{\vec{q}^{\ 4}}{4}\right]\right]\,F^2\\&\times\frac{1}{p^2-m^2_{D^*}+i\epsilon}\,\frac{1}{(p-p_1'+p_2)^2-m_D^2+i\epsilon}\,\frac{1}{(p-p_1)^2-m^2_{\pi}+i\epsilon}\\&\times\frac{1}{(p-p_1')^2-m^2_{\pi}+i\epsilon}\ .
\label{eq:ampl_norel}
\end{split}
\end{equation}
Performing the analytical integration in $dp^0$, we obtain
\begin{equation}
\begin{split}
t&=-\frac{5}{4}g^4\,\frac{1}{2}\,\vec{\epsilon}\ '\ \vec{\epsilon}\ ''\int\frac{d^3p}{(2\pi)^3}\, (\vec{p}^{\ 2}-\vec{q}^{\ 2})\left[(4\vec{p}^{\ 2}-\frac{\vec{q}^{\ 2}}{4})-\frac{1}{\vec{q}^{\ 2}}\left[(2\vec{p}\,\vec{q}\,)^2-\frac{\vec{q}^{\ 4}}{4}\right]\right]\,\frac{F^2}{\omega_1+\omega_2}\,\frac{1}{2\omega_1\omega_2}\\&\times\frac{1}{2E_D}\,\frac{1}{2E_V}[\omega_1^2+\omega_2^2+\omega_1\omega_2-(\omega_1+\omega_2)(2p_1^0-E_V-E_D)+(p_1^0-E_V)(p_1^0-E_D)]\\&\times\frac{1}{p_1^0-\omega_1-E_V+i\epsilon}\,\frac{1}{p_1^0-\omega_1-E_D+i\epsilon}\,\frac{1}{p_1^0-\omega_2-E_V+i\epsilon}\,\frac{1}{p_1^0-\omega_2-E_D+i\epsilon}\ .
\label{eq:ampl_norel2}
\end{split}
\end{equation}

\begin{figure}[htpb]
\centering
\includegraphics[scale=0.3]{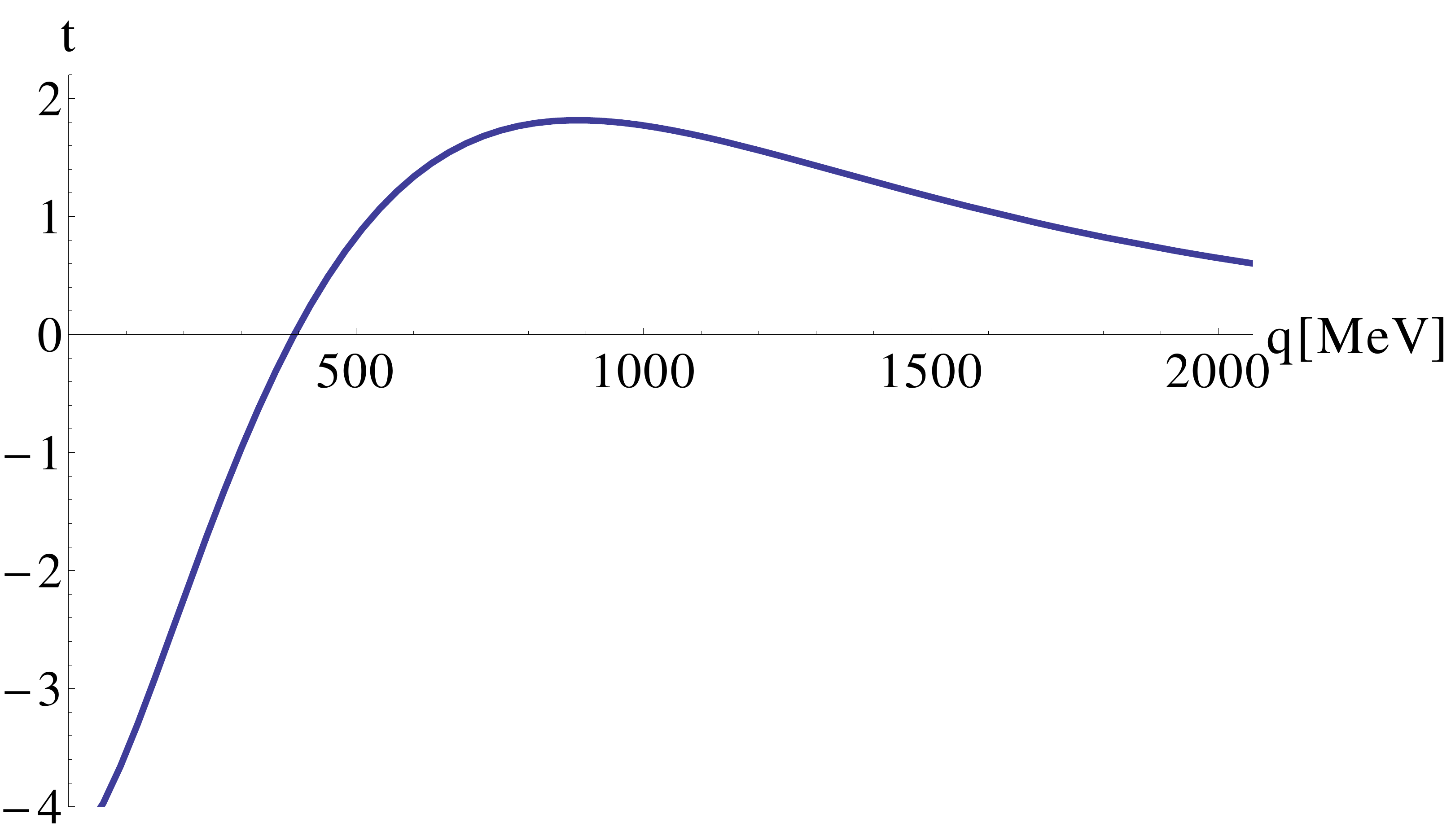}
\caption{Potential $t$ for non-interacting pion exchange as a function of the momentum transferred in the process.}
\label{fig:pionex}
\end{figure}

The potential $t$ is plotted in Fig. \ref{fig:pionex} as a function of the exchanged momentum

Once again, the vector exchange potential in Fig. \ref{fig:vectorq} is dominant in comparison with the contribution of the uncorrelated pion exchange term in Fig. \ref{fig:pionex}.

%**************************************************************************************************
\section{Determination of the $D\bar D^*$ invariant mass distribution for the process $e^+ e^- \to \pi^{\pm} (D \bar D^*)^\mp$}\label{fit}

In Ref.~\cite{Ablikim:2013xfr} the $e^+ e^- \to \pi^{\pm} (D \bar D^*)^\mp$ reaction is studied for a center of mass energy $\sqrt{s}=4.26$ GeV and the $D\bar D^*$ invariant mass associated with this reaction
is obtained, showing a signal around 3885 MeV with a width close to 30 MeV and which is interpreted as a $J^P=1^+$ resonant state. Following Ref.~\cite{alberdd},
we can calculate the $D\bar D^*$ invariant mass spectrum for the reaction studied in Ref.~\cite{Ablikim:2013xfr} as
\begin{align}
\frac{d\sigma}{d M_{D^*\bar{D}^*}}\propto \frac{p\tilde{q}}{s\sqrt{s}}\left|T\right|^2F_L,\label{ss}
\end{align}
where $\sqrt{s}$ is fixed to the value 4.26 GeV, $p$ is the pion momentum in the $e^+ e^-$ center of mass frame, and $\tilde{q}$ is the center of mass momentum in the $D\bar{D}^*$ system:
\begin{align}
p&=\frac{\lambda^{1/2}(s,m^2_\pi,M^2_{D\bar{D}^*})}{2\sqrt{s}},\\
\tilde{q}&=\frac{\lambda^{1/2}(M^2_{D\bar{D}^*},m^2_{D},m^2_{\bar D^*})}{2M_{D\bar{D}^*}}.
\end{align}
The factor $F_L=p^{2L}$ in Eq.~(\ref{ss}) is needed to account for the relative partial wave between the pion and the $D\bar{D}^*$ system produced in the reaction. In this case, we are going to consider the formation of a $J^P=1^+$ state near threshold, thus the $D\bar D^*$ system is preferably produced in S-wave ($L=0$). If a state with mass $M_R$ and width  $\Gamma_R$ is formed in the $D \bar D^*$ system, the amplitude $T$ of Eq.~(\ref{ss}) can be parametrized as
\begin{align}
T=\frac{A}{M^2_{D\bar{D}^*}-M^2_R+iM_R\Gamma_R},\quad A\equiv\textrm{constant}\label{tm}
\end{align}

In general, the $D\bar D^*$ invariant mass distribution can have contributions from a non resonant background. Following Ref.~\cite{Ablikim:2013xfr} we consider a background of the form
\begin{align}
B=\alpha (M_{D\bar{D}^*}-M^\textrm{min}_{D\bar{D}^*})^\beta (M^\textrm{max}_{D\bar{D}^*}-M_{D\bar{D}^*})^\eta\ ,\label{bk}
\end{align}
where $M^\textrm{min}_{D\bar{D}^*}$ and $M^\textrm{max}_{D\bar{D}^*}$ represent the minimum and maximum values of the $D\bar{D}^*$ invariant mass and $\alpha$, $\beta$ and $\eta$ are unknown constants.

In this way, the $D\bar D^*$ invariant mass spectrum can be obtained as
\begin{align}
\frac{d\sigma}{d M_{D\bar{D}^*}}=\frac{1}{s\sqrt{s}}p\tilde{q}\left(\left|T\right|^2F_L+B\right).\label{cross}
\end{align}

As can be seen from Eqs.~(\ref{tm}) and (\ref{bk}), we have 6 unknown parameters to determine the $D\bar D^*$ spectrum (same number as in Ref.~\cite{Ablikim:2013xfr}): the magnitude of the resonant amplitude $A$, the mass and width of the state ($M_R$ and $\Gamma_R$, respectively), the magnitude of the background amplitude, $\alpha$, and the exponents $\beta$ and $\eta$. To constraint these parameters we perform a fit to the data minimizing the $\chi^2$ and consider a value of
the $\chi^2$ per degrees of freedom (d.o.f) around 1 as the criteria to establish the goodness of the fit. This is the same criteria as the one adopted by the authors in Ref.~\cite{Ablikim:2013xfr}, in which a value of $\chi^2/\textrm{d.o.f}$ of 1 is found for the $D^0\bar {D}^{*-}$ mass spectrum and of 1.1 for the $D^+\bar{D}^{*0}$ case.

%**************************************************************************************************
\section{Results}
\subsection{Resonance generation in the $D\bar{D}^*$ system}\label{dd}
Following the scheme of Ref. \cite{melafran}, we roughly compare the strength of the potential in the three cases evaluating $\int V(q) d^3q$. Summing the contributions given by one meson exchange and two pion exchange, with and without interaction, we obtain $\int V(q) d^3q\simeq -112$ GeV$^3$. In the case of vector exchange, the strength is $\int V(q) d^3q\simeq -433$ GeV$^3$. We thus neglect the pseudoscalar exchange contributions but keep them in mind when evaluating uncertainties. 

We studied the $T$ matrix coming from vector exchange for values of $\sqrt{s}$ around $3900$ MeV, in particular the shape of $|T|^2$. 

\begin{figure}[h!]
\centering
\includegraphics[scale=0.4]{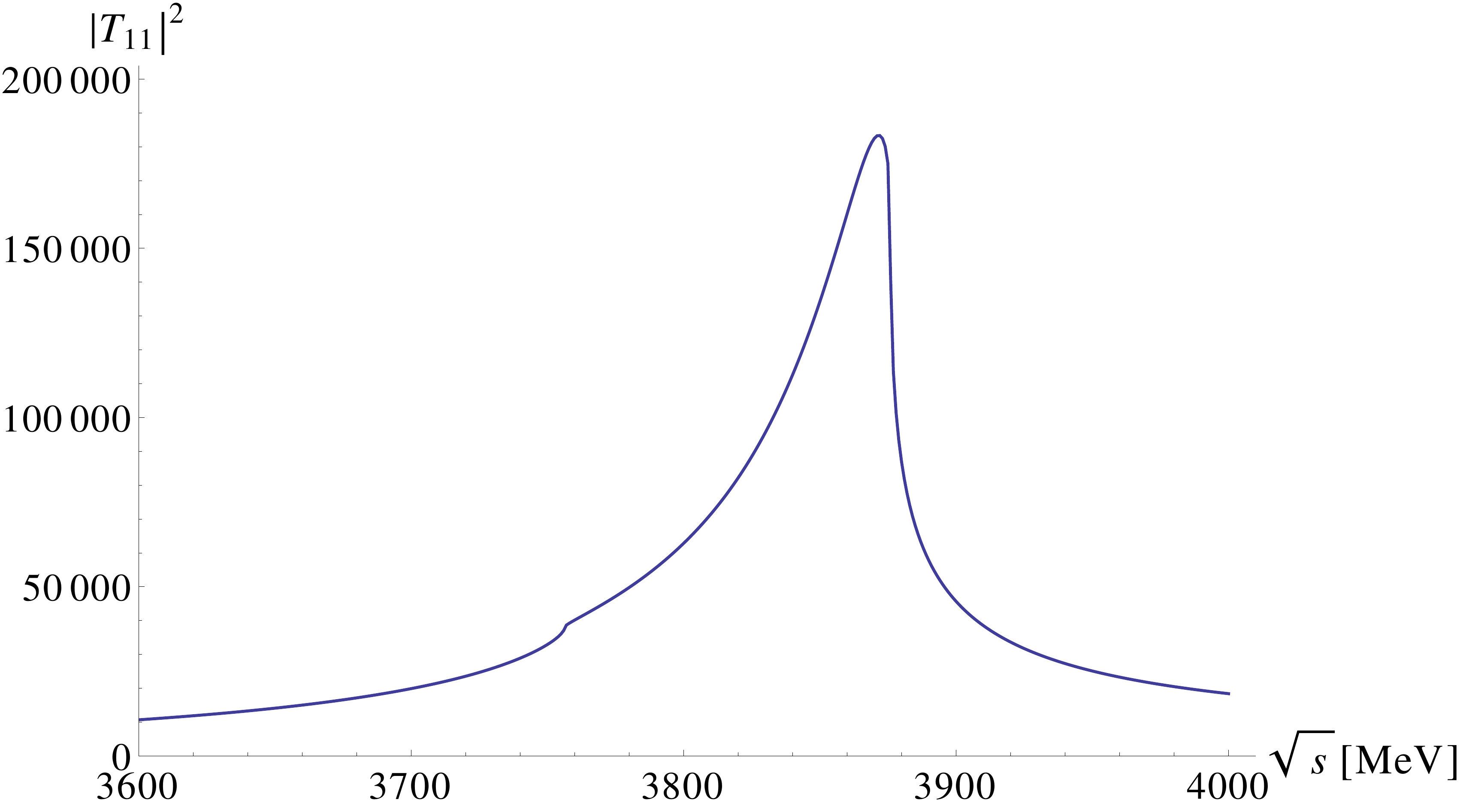}
\caption{$|T|^2$ as a function of $\sqrt{s}$.}
\label{fig:realpole}
\end{figure}

Although no bound state showed up in the $1^-(1^{++})$ case in the region of interest, we found interesting results in the case with positive $G$-parity. In Fig. \ref{fig:realpole}, $|T_{11}|^2$ (where the subscript $11$ means that we are considering the $D\bar{D}^*\rightarrow D\bar{D}^*$ transition), for the case $1^+(1^{+-})$, is  shown as a function of the centre of mass energy. We used the dimensional regularization expression of Eq. \eqref{eq:loopexdm} for the $G$ function, using for the subtraction constants $\alpha_1=-1.28$, $\alpha_2=-1.57$ and $\alpha_3=-1.86$ and choosing $\mu=1500$ MeV, as suggested in \cite{daniel}. This choice of the parameters is equivalent to using a cutoff $q_{max}=770$ MeV.  A clear peak is visible in Fig. \ref{fig:realpole} for $\sqrt{s}=3872$ MeV, with a width of approximately $\Gamma\simeq 40$ MeV. 

\begin{figure}[h!]
\centering
\includegraphics[scale=0.4]{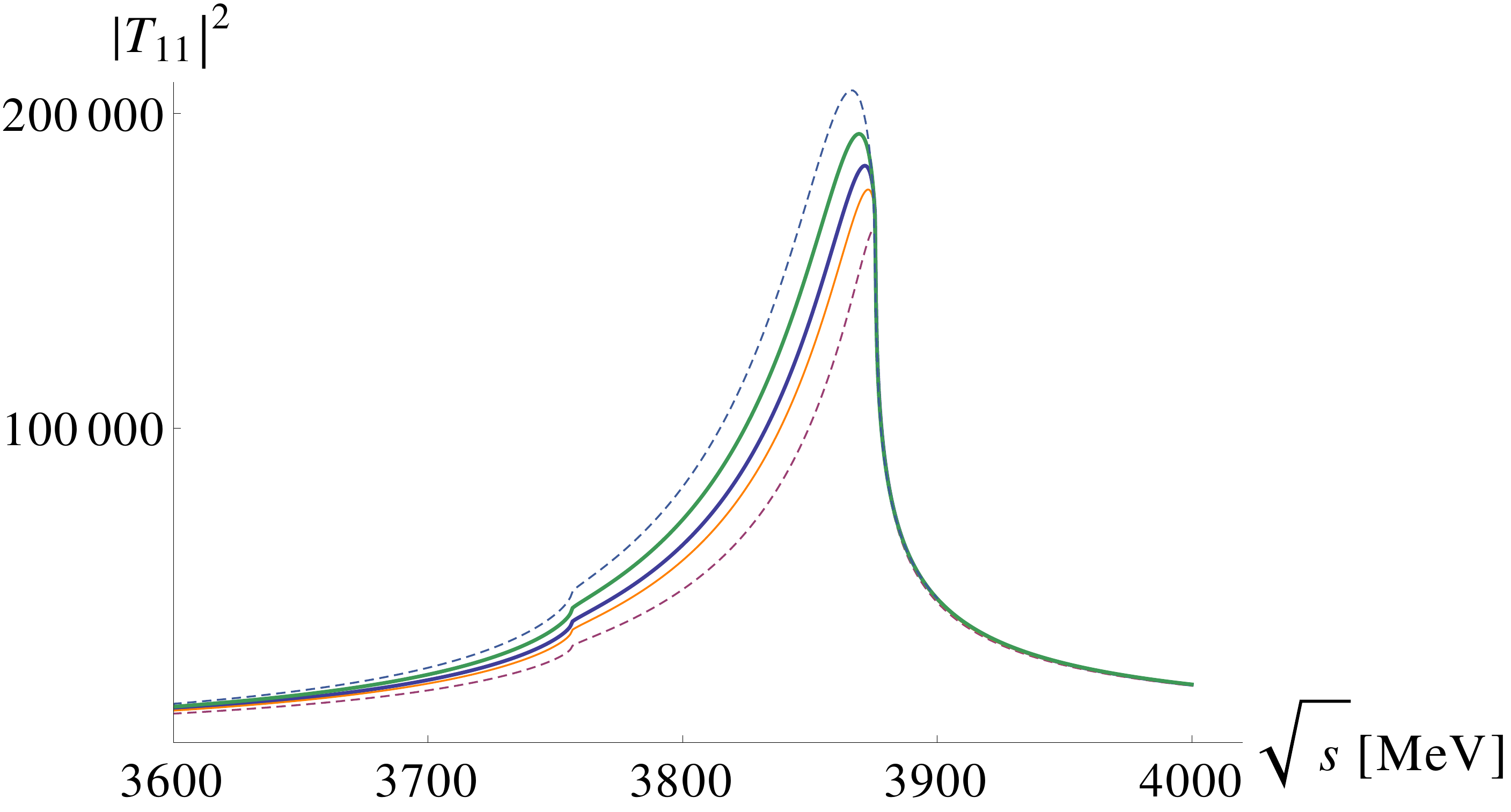}
\caption{$|T|^2$ as a function of $\sqrt{s}$ for values of the cutoff $q_{max}$ equal to $850$, $800$, $770$, $750$ and $700$ MeV. The peak moves to the left as the cutoff increases.}
\label{fig:cutoffs}
\end{figure}

In Fig. \ref{fig:cutoffs} we show the dependence of the position of the peak on the cutoff. The quantity $|T_{11}|^2$ is plotted as a function of $\sqrt{s}$ for values of the $\alpha_i$ subtraction constants corresponding to a cutoff equal to $700$, $750$, $770$, $800$ and $850$ MeV. The corresponding values of the peak are shown in Tab. \ref{tab:peaks}: going to higher values of the cutoff, the binding energy of the state increases. The width varies within $40-50$ MeV. These changes can serve to quantify our uncertainties from the neglected pseudoscalar exchanges or other possible sources. We have also changed the parameter $\Lambda$ in the form factor of Eq. \eqref{eq:ff} in the range $700-1200$ MeV. We have checked that multiplying our potential by a factor within the range of $0.6-1.4$ gives us similar results as with this change of the cutoff and $\Lambda$. The calculations are done using average values of the masses the $D$ and $\bar{D}^*$. If we use the actual masses in the experiments quoted, the changes in the binding energy are of the order of $1$ MeV.

It is interesting to note that the energies obtained all stick around threshold ($3076$ MeV). Next we discuss if there are poles associated to the peaks observed in Fig. \ref{fig:cutoffs}.

\begin{figure}[h!]
\centering
\includegraphics[scale=0.4]{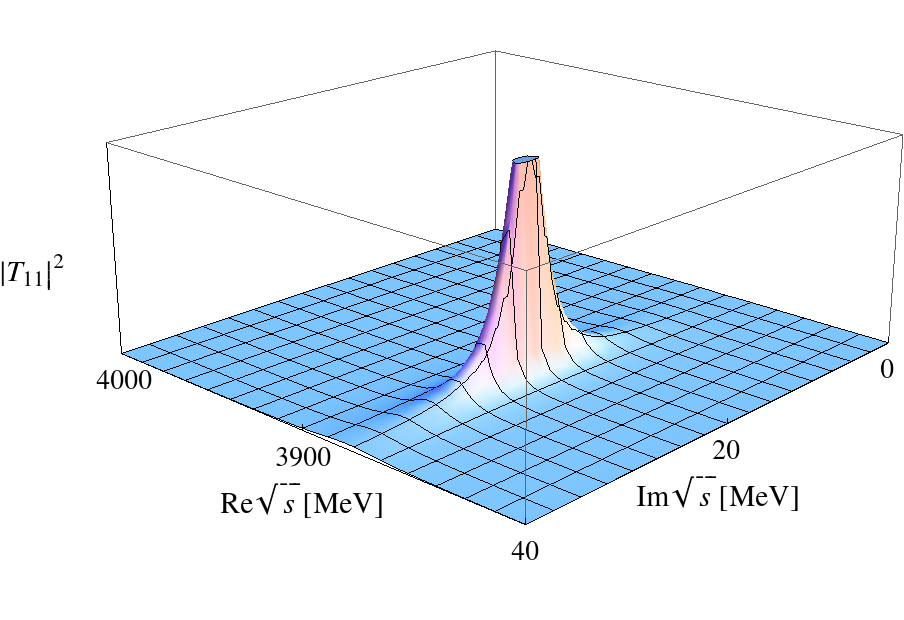}
\caption{$|T|^2$ in the second Riemann sheet for the transition $D\bar{D}^*\rightarrow D\bar{D}^*$ for the $I^G(J^{PC})=1^+(1^{+-})$ sector.}
\label{fig:pole}
\end{figure}

We move to the complex plane, extrapolating the amplitude to complex values of the energy. To do this, for the channels which are open, we need the expression of the loop function in the second Riemann sheet, which can be written as \cite{luisaxial}
\begin{equation}
\label{eq:secondr}
G_i^{II}(\sqrt{s})=G^{I}_i(\sqrt{s})+i\frac{p}{4\pi\sqrt{s}}\ \ \ \ \ \ \ \ \ \ \ Im(p)>0\ ,
\end{equation}
where $G_i^{I}(\sqrt{s})$ is given by Eq. \eqref{eq:loopexdm}.
In Fig. \ref{fig:pole} $|T_{11}|^2$ is plotted in the second Riemann sheet for the value of $q_{max}=770$ MeV. A pole, corresponding to a state with $(\sqrt{s}+i\Gamma/2)=(3878+i23)$ MeV is perfectly visible.

If we lower the cutoff, for a while one still has poles in the complex plane, but for values of $q_{max}<700$ MeV, the poles in $\sqrt{s}$ fade away although one still has a pronounced cusp effect of the amplitude, with experimental consequences in cross sections. This situation is usually referred as having a virtual pole. 

Note that in all cases our states produce peaks around the $D\bar{D}^*$ threshold of $3876$ MeV.

\begin{table}[ tp ]%
\begin{tabular}{c|c}
\hline %
\ \ \ \ $q_{max}\ [\textrm{MeV}]$\ \ \ \  &\ \ \ \ $\sqrt{s}\ [\textrm{MeV}]$\ \ \ \  \\\toprule %
$700$ & $3875$ \\
$750$ & $3873$ \\
$770$ & $3872$ \\
$800$ & $3869$ \\
$850$ & $3867$ \\
\hline
\end{tabular}
\caption{Position of the peak of $|T|^2$ corresponding to different values of $q_{max}$.}
\label{tab:peaks}\centering %
\end{table}

%********************************************************************************************************************************************

\subsection{The $D\bar{D}^*$ invariant mass distribution}
As we have seen in the previous section, the dynamics involved in the $D\bar D^*$ system gives rise to the generation of a state with isospin 1, quantum numbers $J^P=1^+$,  mass $3867-3875$ MeV and width around $40$ MeV. The question which arises now is if a state below the $D\bar D^*$ threshold can be responsible for the signal reported in the $D\bar D^*$ spectrum when studying the reaction $e^+ e^- \to \pi^{\pm} (D \bar D^*)^\mp$ \cite{Ablikim:2013xfr}. 

\begin{figure}[htpb]
\centering
\includegraphics[width=0.45\textwidth]{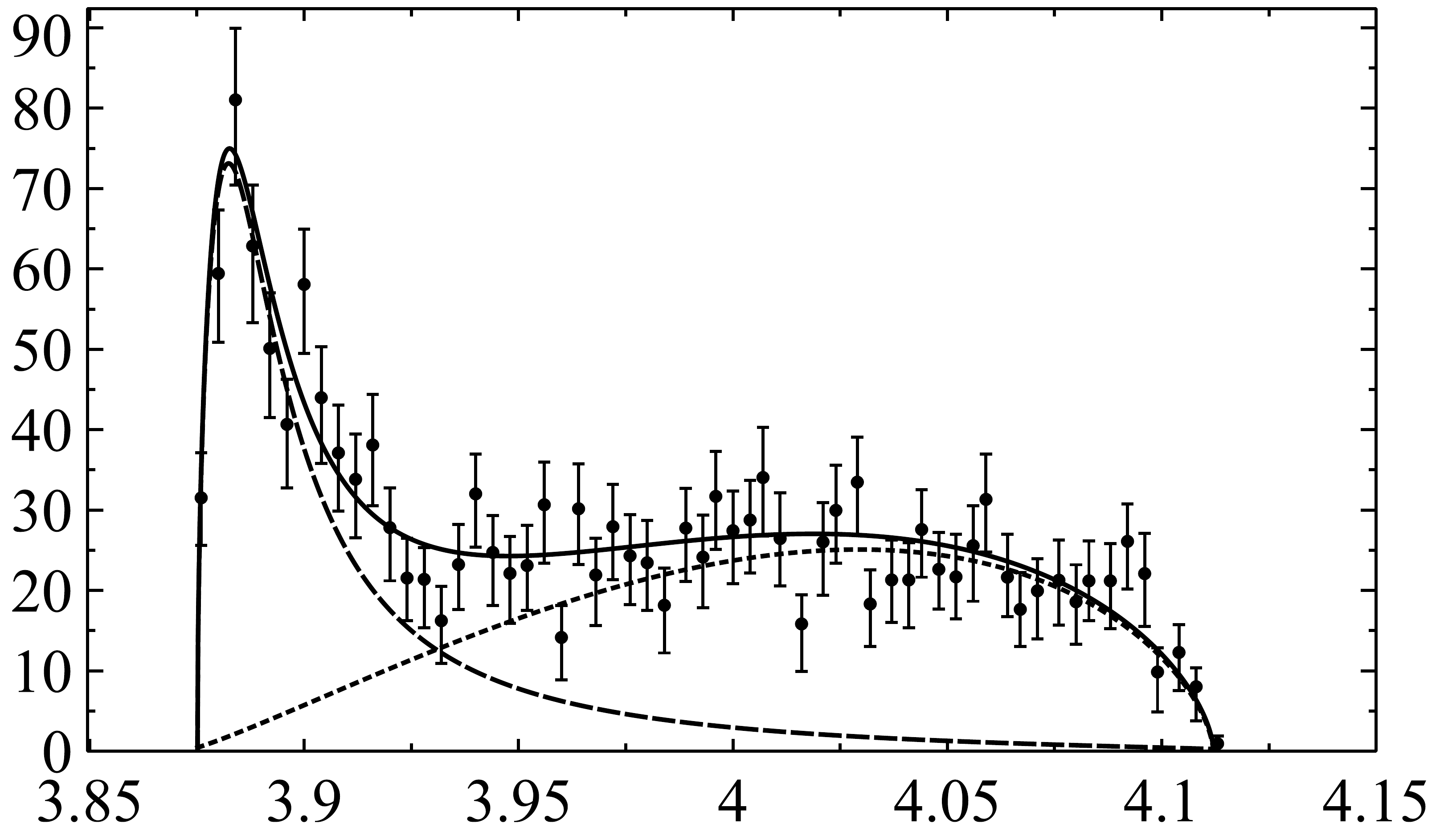}
\includegraphics[width=0.45\textwidth]{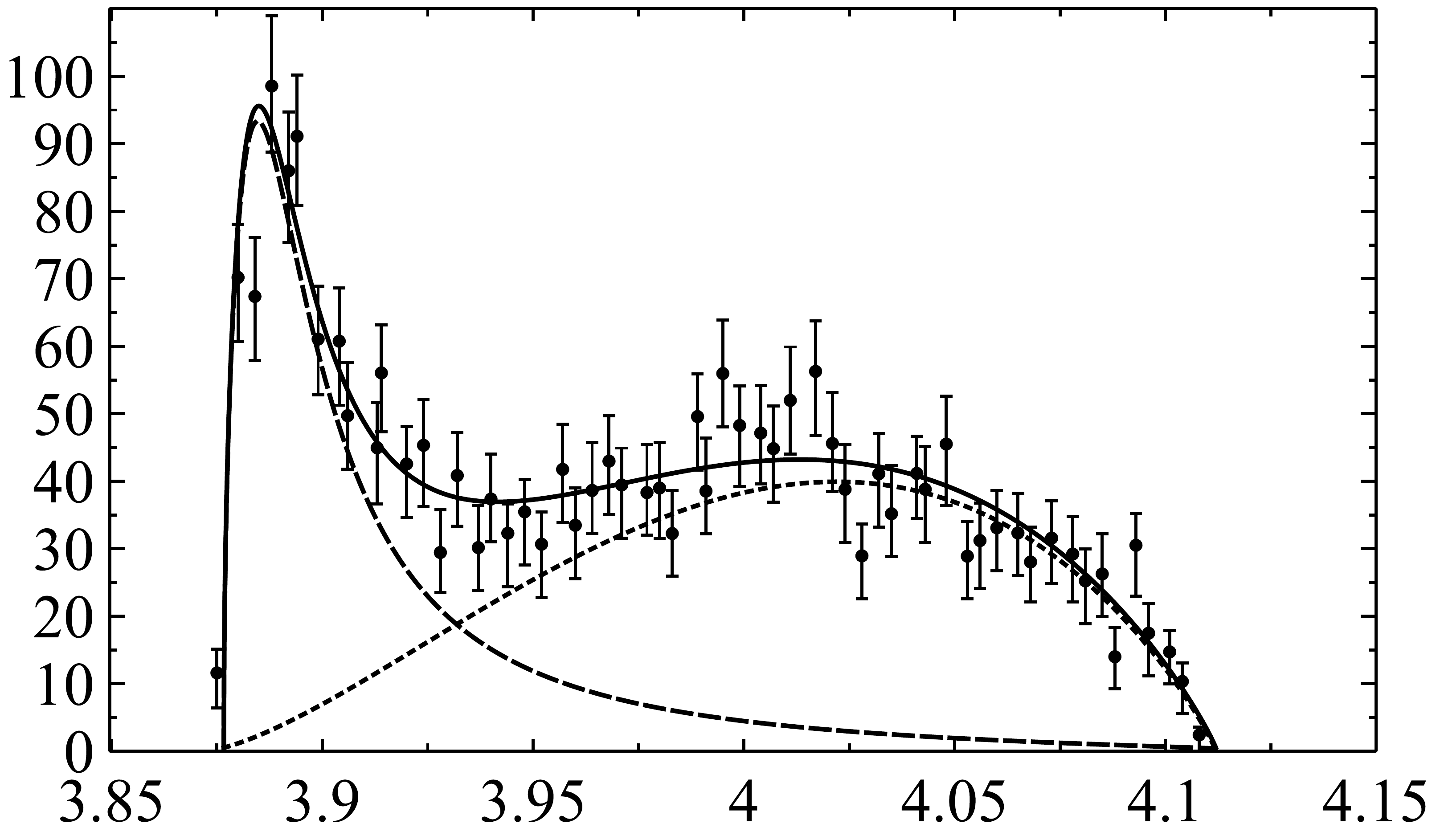}
\caption{Invariant mass distribution for the $D^0D^{*-}$ (left panel) and $D^+\bar{D}^{*0}$ (right panel) systems. The abscissa axis represents the corresponding $D\bar D^*$ invariant mass in units of GeV and  the ordinate axis the spectrum in arbitrary units. The dashed line represents the bound state contribution: $M_R=3874.15$ MeV, width $\Gamma_R=27$ MeV (left panel) and $M_R=3875.62$ MeV, width $\Gamma_R=30$ MeV (right panel). The dotted line corresponds to the background and the solid line is the final result from the fit: $\chi^2/\textrm{d.o.f}=1.3$ (left panel) and $\chi^2/\textrm{d.o.f}=1.1$ (right panel).}
\label{invmassfig}
\end{figure}

Using Eq.~(\ref{cross}) and the procedure explained in Sec.~\ref{fit}, we show in Fig.~\ref{invmassfig} the results found for the $D^0 D^{* -}$ (left panel) and $D^+ \bar {D}^{*0}$ spectra (right panel), respectively, determined considering the formation of a state as the one obtained in our study of the $D\bar D^*$ system. As can be seen, the data can be perfectly explained with a state with a mass close to 3870 MeV and around 30 MeV of width.

We have studied the range of masses that the fit can accommodate. We can have higher masses than $3870$ MeV with still good values of the $\chi^{2}$, but they gradually increase as the mass increases. We put the limit at $3884$ MeV where the $\chi^{2}$ values are no longer good. This gives a range $3862-3884$ MeV, by means of which we can give an acceptable fit to the data. The theory band of $3867-3875$ MeV given in the other section is within the band allowed by the fit to the data.

%******************************************************************************************************************************
\section{Conclusions}

We have done  a combined study of a $Z_c$ state of $I=1$ around $3900$ MeV, which has been claimed in several experiments. On the one hand, we have used an extension of local hidden gauge approach to the heavy quark sector to study the interaction of $D\bar{D}^*$ and $\bar{D}D^*$, together with coupled channels with a pseudoscalar and a vector meson. The constraints of heavy quark spin symmetry show that the terms which are dominant in other processes, like in $I=0$, due to the exchange of light mesons, are now forbidden. Hence, one resorts to  sub-dominant terms that come from the exchange of heavy vectors, or the exchange of two pions. We find that the exchange of two pions is quite small in comparison with the exchange of heavy vectors and its effect is included in the uncertainties of the results. We find a state with a mass of $3869-3875$ MeV and a width around $40$ MeV with $I=1$ and positive $G-$parity. This state, in our formalism, is an isospin partner of the $X(3872)$.

The second part of the work consists in a reanalysis of the experiment of \cite{Ablikim:2013xfr} in the $e^+e^-\rightarrow \pi^{\pm}(D\bar{D}^*)^{\mp}$ reaction. The experimentalists extracted a mass of about $3885$ MeV and width $25\pm 3\pm 11$ MeV from the enhancement of the $D\bar{D}^*$ distribution around threshold. We performed a reanalysis of the data and found a solution close by, with $M_R\simeq 3875$ MeV and $\Gamma\simeq 30$ MeV preferably. Hence, the present study shows that the data of \cite{Ablikim:2013xfr} are compatible with a slightly lower mass, as obtained theoretically in the present paper.

Thus, the results reported here offer a natural explanation of the state claimed in \cite{Ablikim:2013xfr}, in terms of a $D\bar{D}^*(\bar{D}D^*)$ weakly bound state that decays into the $\eta_c\rho$ and $\pi J/\psi$ channels.

The question remains whether that state reconfirmed in this paper would be the same as the $Z_c(3900)$ claimed by BESIII in \cite{Ablikim:2013mio}, or the $Z_c(3894)$ reported by Belle \cite{Liu:2013dau}, or the $Z_c(3886)$ reported by CLEO in  \cite{Xiao:2013iha}. Given the uncertainties in the masses and widths in all these experiments, it is quite likely that these experiments are seeing the same state, although other options cannot be ruled out at the present time. In any case, we can say that, given the fact that a single channel $D\bar{D}^*$ with an energy independent potential cannot produce a resonance above the threshold at $3875.87$ MeV \cite{yama}, a state with $3900$ MeV could not be easily interpreted as a $D\bar{D}^*(\bar{D}D^*)$ molecular state, while the one at lower energy stands naturally for a molecular interpretation, as we have reported here. Further precise measurements and investigations of other decay channels will help shed light on this issue in the future, and they should be encouraged.

%*************************************************************************************************
\section*{Acknowledgments}
This work is partly supported by the Spanish Ministerio de Economia y Competitividad and European FEDER funds under the contract number
FIS2011-28853-C02-01, and the Generalitat Valenciana in the program Prometeo, 2009/090. We acknowledge the support of the European Community-Research Infrastructure Integrating Activity Study of Strongly Interacting Matter (acronym HadronPhysics3, Grant Agreement
n. 283286) under the Seventh Framework Programme of EU. The authors would like to thank the Brazilian funding agencies FAPESP and CNPq for the financial support.

\end{document}